\documentclass[aps,pre,9pt,twocolumn,superscriptaddress,floatfix,showkeys]{revtex4-2}
\pdfoutput=1

\usepackage{amsmath}
\usepackage[dvipsnames]{xcolor}
\usepackage{hyperref}
\newcommand\myshade{85}
\colorlet{mylinkcolor}{violet}
\colorlet{mycitecolor}{YellowOrange}
\colorlet{myurlcolor}{Aquamarine}

\hypersetup{
  linkcolor  = mylinkcolor!\myshade!black,
  citecolor  = mycitecolor!\myshade!black,
  urlcolor   = myurlcolor!\myshade!black,
  colorlinks = true,
}
\usepackage{multirow}
\usepackage{colortbl}
\usepackage{booktabs}
\usepackage{etoolbox}
\usepackage[dvipsnames]{xcolor}
\definecolor{rossoPantano}{RGB}{155,0,20} % padua red
\definecolor{grigioPantano}{RGB}{72,79,89}  % padua gray 
\usepackage{pgfplots}
\usepackage{tikz}
\usepackage{pgfplotstable}
\usetikzlibrary{pgfplots.groupplots,arrows,automata,positioning}
\usepgfplotslibrary{dateplot,fillbetween}
\usepgfplotslibrary{dateplot,colormaps}
\pgfplotsset{compat=newest, lua backend=true}
 \usetikzlibrary {shapes.multipart} 
\usepackage[group-separator={,},group-minimum-digits=4]{siunitx}

\newcommand{\ie}{\textit{i.e.}}

\newenvironment{customlegend}[1][]{%
    \begingroup
    \csname pgfplots@init@cleared@structures\endcsname
    \pgfplotsset{#1}%
}{%
    \csname pgfplots@createlegend\endcsname
    \endgroup
}%

\def\addlegendimage{\csname pgfplots@addlegendimage\endcsname}

\pgfplotsset{
    jitter/.style={
        y filter/.code={\pgfmathparse{\pgfmathresult+rnd*#1}}
    },
    jitter/.default=0.05
}
\pgfdeclarelayer{background}
\pgfdeclarelayer{foreground}
\usepgfplotslibrary{colorbrewer} % L
\pgfsetlayers{background,main,foreground}   %% some additional layers for demo
\AtBeginEnvironment{tabular}{
\sffamily\fontsize{7.5}{10}\selectfont
}
\newcommand{\addtabletext}[1]{{\setlength{\leftskip}{9pt}\fontsize{7}{9}\selectfont#1}}

\tikzstyle{every picture}+=[font=\sffamily] 
\pgfplotsset{every axis/.append style={
    font=\sffamily,
    label style={font=\sffamily\normalsize},
    title style={text depth=0.5ex,font=\sffamily\normalsize},
    tick label style={font=\sffamily\footnotesize, 
        /pgf/number format/assume math mode=true},  % for sans serif tick labels
    xtick align = center, 
    ylabel near ticks,
    y label style={font=\sffamily\normalsize},
    xlabel near ticks,
    x label style={font=\sffamily\normalsize},
    legend cell align={left},
    legend style={draw=none, font=\sffamily\small},
    },
    legend image code/.code={
    \draw[mark repeat=2,mark phase=2]
        plot coordinates {
        (0cm,0cm)
        (0.15cm,0cm)        %% default is (0.3cm,0cm)
        (0.3cm,0cm)         %% default is (0.6cm,0cm)
        };%
    },
    boxplot/hide outliers/.code={
        \def\pgfplotsplothandlerboxplot@outlier{}%
    }
}
\pgfplotsset{compat=newest}  
\pgfplotsset{
    jitter/.style={
        x filter/.code={\pgfmathparse{\pgfmathresult+(rnd-.5)*#1}}
    },
    jitter/.default=0.1
}
\usepackage{cellspace}

\pgfplotstableset{
    /color cells/min/.initial=0,
    /color cells/max/.initial=1000,
    /color cells/textcolor/.initial=,
    color cells/.code={%
        \pgfqkeys{/color cells}{#1}%
        \pgfkeysalso{%
            postproc cell content/.code={%
                \begingroup
                \pgfkeysgetvalue{/pgfplots/table/@preprocessed cell content}\value
                \ifx\value\empty
                    \endgroup
                \else
                    \pgfmathfloatparsenumber{\value}%
                    \pgfmathfloattofixed{\pgfmathresult}%
                    \let\value=\pgfmathresult
                \newdimen\flagdim
                \flagdim = 5 pt
                \newdimen\valdim 
                \valdim = \value pt
                
                \ifdim \flagdim = \valdim %  # check if the value passed matches the flag 
                    \def\pgfmathresult{0.4, 0.4, 0.4} % set pgfmathresult to gray 
                \else 
                    \pgfplotscolormapaccess
                        [\pgfkeysvalueof{/color cells/min}:\pgfkeysvalueof{/color cells/max}]
                        {\value}
                        {\pgfkeysvalueof{/pgfplots/colormap name}}%
                \fi 
                
                \def\typesetvalue{\phantom{XXX }}  % we dont care about the values
                \pgfkeysgetvalue{/color cells/textcolor}\textcolorvalue
                \toks0=\expandafter{\typesetvalue}%
                \xdef\temp{%
                    \noexpand\pgfkeysalso{%
                        @cell content={%
                            \noexpand\cellcolor[rgb]{\pgfmathresult}%
                            \noexpand\definecolor{mapped color}{rgb}{\pgfmathresult}%
                            \ifx\textcolorvalue\empty
                            \else
                                \noexpand\color{\textcolorvalue}%
                            \fi
                            \the\toks0 %
                        }%
                    }%
                }%
                \endgroup
                \temp
                \fi
            }%
        }%
    }
}

\newcolumntype{?}{!{\color{black}\vrule width 0.2pt}}
\newlength{\cellspacelimit}
\arrayrulecolor{black}

\newcommand{\heatmaprow}[2]{\tiny \pgfplotstabletypeset[
    color cells={min=-#1,max=#1, textcolor=black},
    col sep=comma, row sep=crcr,
    header=false,
    fixed, precision=1, fixed zerofill,
    /pgfplots/colormap={custom}{color=(RdBu-L), color=(white),color=(RdBu-B)},
    every head row/.style={output empty row},
    every column/.style={column type={@{}l@{}!{\color{black}\vrule}}},
    every first row/.style={
        before row=\specialrule{0.01em}{0em}{0em},
        after row=\specialrule{0.01em}{0em}{0em},
    },
    every first column/.style={
        column type/.add={|}{},
    },
]{#2}}

\pgfplotsset{
    cycle list/Set1-5,
    cycle multiindex* list={
        mark list*\nextlist
        Set1-5\nextlist
    }
}

\usetikzlibrary{automata,positioning,arrows,calc}
\usepackage{tikzpeople}

\tikzstyle{edge} = [thick]  % edge style
\tikzstyle{diredge} = [edge, draw=Set3-D, very thick, -{Stealth[Set3-D, fill=Set3-D]}]
\tikzstyle{faded} = [draw opacity=0.3, fill opacity=0.3]

\tikzstyle{a0} = [alice, label={[yshift=5pt, font=\footnotesize]$a_0$}]
\tikzstyle{a1} = [bob, shirt=red, label={[yshift=5pt, font=\footnotesize]$a_1$}]
\tikzstyle{a2} = [charlie, female, label={[yshift=5pt, font=\footnotesize]$a_2$}]
\tikzstyle{a3} = [dave, label={[yshift=5pt, font=\footnotesize]$a_3$}]
\tikzstyle{a4} = [person, hair=RedOrange, shirt=violet, label={[yshift=5pt, font=\footnotesize]$a_4$}]
\tikzstyle{a5} = [bob, female, label={[yshift=5pt, font=\footnotesize]$a_5$}]
\tikzstyle{a6} = [alice, skin=Tan, shirt=RoyalBlue, details=Red, label={[yshift=5pt, font=\footnotesize]$a_6$}]

\tikzstyle{a0_} = [alice, outer sep=20pt]
\tikzstyle{a1_} = [bob, shirt=red, outer sep=20pt]
\tikzstyle{a2_} = [charlie, female, outer sep=20pt]
\tikzstyle{a3_} = [dave, outer sep=20pt]
\tikzstyle{a4_} = [person, hair=RedOrange, shirt=violet, outer sep=20pt]
\tikzstyle{a5_} = [bob, female, outer sep=20pt]
\tikzstyle{a6_} = [alice, skin=Tan, shirt=RoyalBlue, details=Red, outer sep=20pt]

\newcommand{\paper}[2]{
	\tikz[scale=1,line width=1pt]{
        \def\a{1.10} % relative height
        \def\b{0.15}  % relative height/width of corner
        \def\c{0.2}  % relative margin width (on either side)
        \def\d{0.05} % vertical offset of lines
        \def\N{3}    % number of lines
        \def\v{2.2} % line spacing
        
        \def\authors{#2}
        \pgfmathparse{dim(\authors)}
        \def\tmp{\pgfmathresult}
        \pgfmathsetmacro\A{\tmp}  % roundabout way of storing length of passed array
        \draw [#1, fill=lightgray!20]        
        (0,0)
                --  ++(-1,0)
                --  ++(0,\a)
                --  ++(1-\b,0)
                --  ++(\b,-\b)
                -- cycle;
        \draw[draw=#1!75, fill=#1] (0,\a-\b) -- ++(-\b,0) -- ++ (0,\b);
        
         \foreach \i [evaluate=\i as \auth using ({\authors[\i-1])}] in {1,...,\A}{%
            \pgfmathsetmacro\xline{\a-\d-\i*\a/(\A+1)}
            \node[\auth, scale=0.68] at (-\xline, 0.75) {};
         }
        
        \foreach \lnum in {1,...,\N}{
           \pgfmathsetmacro\yline{\a-\d-\lnum*\a/(\N+1)}
           \pgfmathsetmacro\yline{\yline/\v}
            \draw [#1!75] (-1+\c,\yline) -- (-\c,\yline);
         }
    }	
}

\usepackage[capitalise]{cleveref}
\Crefname{equation}{Eq.}{Eqs.}
\Crefname{figure}{Fig.}{Figs.}
\Crefname{tabular}{Tab.}{Tabs.}
\usepackage{comment}

\usepackage{orcidlink}

\tikzstyle{active} = [
    smooth, thick,
    draw=Pastel1-A!60!red, 
    mark=*, mark options={draw=Pastel1-A!30!red, thick, fill=Pastel1-A!80, fill opacity=1},
    mark size=1.8, 
    text=black,
    draw opacity=0.8,
    fill opacity=0.8,
]

\tikzstyle{active_err} = [
    fill=Pastel1-A!70,
    fill opacity=0.8,
]

\tikzstyle{active_large} = [
    active, thick,
    mark size=1.5
]

\tikzstyle{active2} = [
    smooth,
    draw=pink!50!red, 
    mark=oplus*, mark size=1.4, mark options={fill=white},
    text=black,
    draw opacity=0.9,
    fill opacity=0.9,
]

\tikzstyle{active3} = [
    active, thick,
    draw=Accent-A!60!black,
    mark=*, mark size=1.8, mark options={fill=Accent-A!50!white, fill opacity=1},
]

\tikzstyle{active3_err} = [
    fill=Accent-A!50,
    fill opacity=0.8,
]

\tikzstyle{baseline} = [
    smooth, thick,
    draw=Pastel1-B!60!blue, mark=square*, 
    mark size=1.65, 
    mark options={fill=white, fill opacity=1},
    text=black,
    draw opacity=0.8,
    fill opacity=0.8,
]

\tikzstyle{baseline_err} = [
    fill=Pastel1-B!60,
]

\tikzstyle{baseline_large} = [
    baseline, thick,
    mark size=1.2,
]

\tikzstyle{baseline2} = [
    smooth,
    draw=cyan!50!black, mark=halfsquare*, 
    mark size=1.4, 
    mark options={draw=cyan!50!black, mark color=cyan, fill=white},
    text=black,
    draw opacity=0.9,
    fill opacity=0.9,
]

\tikzstyle{baseline3} = [
    smooth, thick,
    draw=Accent-B!75!black, mark=square*, mark size=1.8, 
    mark options={fill=Accent-B!60, fill opacity=1},
    text=black,
    draw opacity=0.9,
    fill opacity=0.9,
]

\tikzstyle{baseline3_err} = [
    fill=Accent-B!50,
    fill opacity=0.8,
]

\tikzstyle{cumulative_prob} = [
    smooth, thick,
    Mulberry,
    mark=*, 
    mark options={draw=Mulberry, fill=Mulberry!20, scale=0.5},
    draw opacity=0.8,
    fill opacity=0.8,
]

\tikzstyle{cumulative_err} = [
    fill=Mulberry!50,
    fill opacity=0.5,
]

\tikzstyle{cumulative_baseline} = [
    smooth,
    JungleGreen,
    thick, mark=none, 
    draw opacity=0.8,
]

\tikzstyle{cumulative_baseline_err} = [
    fill=JungleGreen!40,
    fill opacity=0.4,
]

\tikzstyle{cumulative_baseline2} = [
    smooth, 
    YellowGreen, thick, mark=none, 
    draw opacity=0.8,
]

\tikzstyle{cumulative_baseline_err2} = [
    fill=YellowGreen!40,
    fill opacity=0.6,
]

\tikzstyle{inactive} = [
    smooth, thick,
    draw=Set2-F!60!brown, mark=pentagon*, 
    mark size=1.9, 
    mark options={fill=Set2-F!60},
    text=black,
    draw opacity=0.8,
    fill opacity=0.8,
]

\tikzstyle{inactive_err} = [
    fill=Set2-F!80
]

\tikzstyle{error1} = [Accent-A, draw opacity=0.8, only marks, mark options={draw=Accent-A!70!black, thick, fill=Accent-A!70!white}, yshift=-3pt, mark size=1.3]
\tikzstyle{error2} = [PiYG-E, mark options={draw=PiYG-E!70!black, thick, fill=PiYG-E!70!white}, draw opacity=0.8, only marks, yshift=-8pt, mark size=1.3]
\tikzstyle{error3} = [Accent-C, draw opacity=0.8, only marks, yshift=-5pt]

\pgfplotscolorbarCMYKworkaroundfalse

\providecommand\JournalTitle[1]{#1}

\begin{document}
\title{Collaboration and topic switches in science}

% Use letters for affiliations, numbers to show equal authorship (if applicable) and to indicate the corresponding author
\author{Sara Venturini\,\orcidlink{0000-0002-2653-8533}}
\thanks{SV and SS contributed equally to this work.}
\affiliation{Department of Mathematics “Tullio Levi-Civita”, University of Padova, Padova 35121, Italy}

\author{Satyaki Sikdar\,\orcidlink{0000-0003-1669-6594}}
\thanks{SV and SS contributed equally to this work.}
\affiliation{Luddy School of Informatics, Computing, and Engineering, Indiana University, Bloomington, IN 47408, USA}

\author{Francesco Rinaldi\,\orcidlink{0000-0001-8978-6027}} 
\affiliation{Department of Mathematics “Tullio Levi-Civita”, University of Padova, Padova 35121, Italy}

\author{Francesco Tudisco\,\orcidlink{0000-0002-8150-4475}}
\affiliation{School of Mathematics, Gran Sasso Science Institute, L'Aquila 67100, Italy}

\author{Santo Fortunato\,\orcidlink{0000-0002-9039-4730}}
\email[Corresponding author: ]{santo@indiana.edu}
\affiliation{Luddy School of Informatics, Computing, and Engineering, Indiana University, Bloomington, IN 47408, USA}
\affiliation{Indiana University Network Science Institute (IUNI), Bloomington, IN 47408, USA}


\begin{abstract}
    Collaboration is a key driver of science and innovation. Mainly motivated by the need to leverage different capacities and expertise to solve a scientific problem, collaboration is also an excellent source of information about the future behavior of scholars. 
    In particular, it allows us to infer the likelihood that scientists choose future research directions via the intertwined mechanisms of selection and social influence. 
    Here we thoroughly investigate the interplay between collaboration and topic switches. 
    We find that the probability for a scholar to start working on a new topic increases with the number of previous collaborators, with a pattern showing that the effects of individual collaborators are not independent. 
    The higher the productivity and the impact of authors, the more likely their coworkers will start working on new topics. 
    The average number of coauthors per paper is also inversely related to the topic switch probability, suggesting a dilution of this effect as the number of collaborators increases.
\end{abstract}

% \date{\today}
% \date{}

\maketitle

Modern science has become increasingly collaborative over the past decades~\cite{wuchty07}. Large teams have become almost necessary to tackle complex problems in various disciplines, requiring a large pool of knowledge and skills. On the other hand, small teams may introduce novel paradigms~\cite{wu19}. 

A powerful representation of the collaborative nature of science is given by a collaboration network, in which nodes are authors, and two nodes are connected if they have coauthored at least one paper.
With the growing availability of bibliometric data, collaboration networks have been extensively studied, and their structural properties are now well known~\cite{newman01,guimera05,pan12,petersen15}.

Collaboration networks are concrete manifestations of \textit{homophily} between scholars. People working on the same topic or problem may decide to team up and leverage their respective skills to increase their chances of discovering new results. This is an example of \textit{selection}, in that similar individuals end up interacting with each other.

On the other hand, collaboration could also induce \textit{social influence}, in that scholars might affect the future behavior of their coauthors. Coauthors often expose us to new tools, methods, and theories, even when the latter is not being used for the specific project carried out by the team. The link between diffusion of knowledge and collaboration has been highlighted and explored for some time. For instance, it is known that knowledge flow occurs with a greater probability between scholars who have collaborated in the past~\cite{singh05} and those who are in close proximity in the network~\cite{sorenson04}. 

In particular, once scholars discover new research topics, they may decide to work on them in the future. Switches between research interests have become increasingly frequent over time~\cite{zeng19} and have been quantitatively investigated~\cite{jia17,zeng22}.
The decision to switch may actually be induced by the coauthors in a social contagion process~\cite{centola07, christakis07,leskovec07,centola10,bond12} where scholar~$a$, who spreads the new topic, influences scholar $b$ to adopt it. 
For this reason, epidemic models have been applied to describe the diffusion of ideas~\cite{goffman64,goffman66,bettencourt06}. In these models, an \textit{infected} individual $a$ exposes a \textit{susceptible} individual $b$ to a disease with a certain probability of getting infected and continuing the spread. In the case of a topic, the infection spreads if $b$ works on the new topic. 

Here we present an extensive empirical analysis of the relationship between topic switches of scientists and their collaboration patterns. 
%We will build upon the analysis of evolving social networks and contexts presented in Ref.~\cite{kossinets06}. 
We distinguish active authors, \ie, those who have papers on the new topic, from inactive authors who have never published in that area.
For simplicity, we focus only on the first-order neighborhoods in the collaboration network.
We find that the probability for an inactive scholar to switch topic grows with the productivity and impact of their active coauthors. The larger the average number of inactive coauthors of active scientists, the smaller the effect. Also, the topic-switch probability for an inactive scholar grows with the number of their active coauthors, with a profile suggesting that the contributions of each coauthor are not independent.

\begin{figure*}[htb]
    \centering
    
    % notes  
% a0 is highly active 
% a1, a2 filler authors 
% a3, a4, a5 never gets active - 
% in observation: a3 writes a paper with a0, a4 gets active with someone else, a5 does not become active 
% a4: has 2 exposures, a5 has 1, add new author a6 who has 0?
% add a list of infected and susceptible authors to the top of the list??

% a0 highly active 
% a6 inactive, 2 contacts with a0, 1 each with a1, a3 

% \tikzsetnextfilename{toy-example-main}

\def\paperauthors{{
    {"a0", "a1"}, {"a0", "a1", "a6"}, {"a0", "a3"}, {"a4", "a5"}, {"a1", "a5", "a6"}, {"a0", "a2", "a3"}, {"a5", "a6"}
    % {"a1", "a3"}, {"a0", "a1", "a2"}, {"a2", "a5"}, {"a0", "a6"}, {"a3", "a4"}, 
    % {"a0", "a4", "a5"}, {"a0", "a1"}, {"a0", "a5", "a6"}, 
    % {"a0", "a1", "a6"}
}}

\def\paperauthorsobs{{
    {"a0", "a4"}, {"a0", "a2"}, {"a4", "a5"}, 
    {"a1", "a5"}, {"a2", "a6"}, {"a1", "a2", "a4"}
}}

\begin{tikzpicture}[help lines, scale=0.9, transform shape]
    
    % \draw[help lines, draw opacity=1] (-1,-2) grid +(11,3);
	\def\ang{45}
    % \node at (5, 0) {Papers: \numpapers \print{\papercolors[0]}};

    \begin{scope}    % panel A
    	\node at (-1.5,1.8) {\large{\textbf{A}}};
        \begin{scope}[shift={(0.2,-0.3388)}]  % exposure window
            \node[rotate around y=\ang] at (-1.05,1.2) {\paper{Set3-D}{\paperauthors[0]}};
            \node[rotate around y=\ang] at (0.05,0.9444) {\paper{gray}{\paperauthors[1]}};
            \node[rotate around y=\ang] at (1.15,0.578) {\paper{gray}{\paperauthors[2]}};
            \node[rotate around y=\ang] at (2.25,0.3326) {\paper{Set3-D}{\paperauthors[3]}};
            \node[rotate around y=\ang] at (3.35,0.0164) {\paper{gray}{\paperauthors[4]}};
            \node[rotate around y=\ang] at (4.45,-0.2392) {\paper{gray}{\paperauthors[5]}};
            \node[rotate around y=\ang] at (5.55,-0.5388) {\paper{gray}{\paperauthors[6]}};
            % \node[rotate around y=\ang] at (6.65,-0.85) {\paper{gray}{\paperauthors[7]}};
            % \node at (10, 1) {\paper{gray}{\paperauthors[5]}};  
        \end{scope}

        %%%%%%%%%%%%%%
        \begin{scope}[shift={(7.65,-0.3388)}]  % observation window
            \node[rotate around y=\ang] at (0.05,0.9444) {\paper{Set3-D}{\paperauthorsobs[0]}};
            \node[rotate around y=\ang] at (1.15,0.6332) {\paper{Set3-D}{\paperauthorsobs[1]}};
            \node[rotate around y=\ang] at (2.25,0.3326) {\paper{Set3-D}{\paperauthorsobs[2]}};
            \node[rotate around y=\ang] at (3.35,0.0164) {\paper{Set3-D}{\paperauthorsobs[3]}};
            \node[rotate around y=\ang] at (4.4604,-0.3172) {\paper{Set3-D}{\paperauthorsobs[4]}};
            \node[rotate around y=\ang] at (5.5708,-0.5952) {\paper{Set3-D}{\paperauthorsobs[5]}};
        \end{scope}
        
        \node at (2.15,1.8) {Interaction Window (IW)};
        \node at (10.25,1.8) {Activation Window (AW)};

        %% line 
        \node (v1) at (-2.5,-0.5) {};
        \node (v2) at (15,-0.5) {};
        \draw[very thick, -stealth]  (v1) edge (v2);

        %% marks on the line
        \node (v3) at (-1.75,-0.2) {};
        \node (v4) at (-1.75,-0.8) {};
        \draw[very thick]  (v3) edge (v4);    
        \node at (-1.75, -1) {$T_0-T$};
        
        \node (v3) at (6.5,-0.2) {};
        \node (v4) at (6.5,-0.8) {};
        \draw[very thick]  (v3) edge (v4);    
        \node at (6.5, -1) {$T_0$};

        \node (v3) at (14,-0.2) {};
        \node (v4) at (14,-0.8) {};
        \draw[very thick]  (v3) edge (v4);    
        \node at (14, -1) {$T_0+T$};
    \end{scope}
    
    %%%%%%%%%%%%%
    \begin{scope}    % panel B
    	\node at (-1.5,-1.6) {\large{\textbf{B}}};
    	% collab graph
    	\begin{scope}[scale=0.9, transform shape, shift={(-0.5915,-1.4444)}]
    		\node[a0_, label={[text=Set3-D]below:$a_0$}] (a0) at (1.75,-2.5) {};
    		\node[a1_, label={[text=Set3-D]above:$a_1$}] (a1) at (1.75,-0.5) {};
    		\node[a2_, label={above:$a_2$}] (a2) at (3.5,-1.5) {};
    		\node[a3_, label={below:$a_3$}] (a3) at (3.5,-3.5) {};
    		\node[a4_, label={[text=Set3-D]below:$a_4$}] (a4) at (1.75,-4.5) {};
    		\node[a5_, label={[text=Set3-D]below:$a_5$}] (a5) at (0,-3.5) {};
    		\node[a6_, label={above:$a_6$}] (a6) at (0,-1.5) {};
    		
    		\draw[edge] (a0) edge[right] node {2} (a1);
     		\draw[edge] (a0) edge[above] node {} (a2);
     	% 	\draw[edge] (a0) edge (a4);
     	% 	\draw[edge] (a0) edge node[above]{2} (a5);
            \draw[edge] (a0) edge node [above] {} (a3);
    		\draw[edge] (a0) edge node [above] {} (a6);
    		
            \draw[edge] (a1) edge (a6);
            \draw[edge] (a1) edge (a5);
            
     		\draw[edge] (a4) edge (a5);
    		\draw[edge] (a2) edge (a3);
            \draw[edge] (a5) edge node [left] {2} (a6);
    	\end{scope}
    	% \node at (0.7928,-7) {Collaboration graph};
    \end{scope}

	\begin{scope}[shift={(4.2776,-1.05)}]  % panel C - exp 1
        \node at (0.6,-0.55) {\large{\textbf{C}}};

        \begin{scope}[scale=0.9, transform shape, shift={(3.1202,-3)}]
            \node[a6_, label={right:$a_6$}] (a6) at (0,0.2468) {};
            \node[text=Set3-D] at (-0.5,0.1234) {(4)};
            
    	    \node[a0_, label={[Set3-D]above:$a_0$}] (a0) at (-1.7436,1.25) {};	
            \node[a1_, label={[Set3-D]above:$a_1$}] (a1) at (1.8,1.25) {};
            \node[a5_, label={[Set3-D]below:$a_5$}] (a5) at (0,-1.75) {};
    		
            \draw[diredge] (a0) edge (a6);
            \draw[diredge] (a1) edge (a6);
            \draw[diredge] (a5) edge[bend left] (a6);
            \draw[diredge] (a5) edge[bend right] (a6);
        \end{scope}
	\end{scope}

%     %%%%%%%%%%%%
	\begin{scope}[shift={(2,0.0545)}]  % panel D - exp 1
		\node at (8.4476,-1.6548) {\large{\textbf{D}}};
		
		\begin{scope}[scale=0.9, transform shape, shift={(10.4287,-1.5)}]  % mark authors who are already active (a1 and a2)
			\node[a0_, label={[text=Set3-D]below:$a_0$}] (a0) at (1.75,-2.5) {};
    		\node[a1_, label={[text opacity=0.2]above:$a_1$}, faded] (a1) at (1.75,-0.5) {};
    		\node[a2_, label={above:$a_2$}] (a2) at (3.5,-1.5) {};
    		\node[a3_, label={below:$a_3$}] (a3) at (3.5,-3.5) {};
    		% \node[a4_, faded, label={[text opacity=0.2]below:$a_4$}] (a4) at (1.75,-4.5) {};
    		% \node[a5_, faded, label={[text opacity=0.2]below:$a_5$}] (a5) at (0,-3.5) {};
    		\node[a6_, faded, label={[text opacity=0.2]above:$a_6$}] (a6) at (0,-1.5) {};

            \draw[diredge, faded] (a0) edge[right] node {2} (a1);
     		\draw[diredge] (a0) edge[above] node {} (a2);
     	% 	\draw[edge] (a0) edge (a4);
     	% 	\draw[edge] (a0) edge node[above]{2} (a5);
            \draw[diredge] (a0) edge node [above] {} (a3);
    		\draw[diredge, faded] (a0) edge node [above] {} (a6);
	    \end{scope}

        % \node at (10.8,-7) {Measuring an author's influence};
	\end{scope}
% %\end{scope}

\end{tikzpicture}
    
    \caption{Schematic setup for our analysis. (A) Stream of papers across interaction (IW) and activation (AW) windows. Papers tagged with the focal topic $t$ are marked in red. (B) Author collaboration graph at the end of IW. Authors $a_i$ and $a_j$ are linked by an edge of weight $k$ if $a_i$ coauthored $k$ papers with $a_j$ within the IW. The authors active in the focal topic by the end of IW are marked in red. (C) 
    Focus: inactive authors. Inactive author $a_6$ has four active contacts from three sources \{$a_0$, $a_1$, $a_5$\} derived from the collaboration graph in (B). (D) Focus: active authors. Active author $a_0$ has four coauthors \{$a_1$, $a_2$, $a_3$, $a_6$\}, of whom $a_1$ is already active, and $a_6$ also collaborated with $a_1$ in the IW. This leaves the subset of exclusive inactive coauthors $\{a_2, a_3\}$. Within this subset, only $a_2$ becomes active in the AW, resulting in $a_0$'s source activation probability of $\tfrac{1}{2}=0.50$. Additionally, $a_2$ writes their first paper with $a_0$ in the AW.}
    
    \label{fig:toy-example}
\end{figure*}
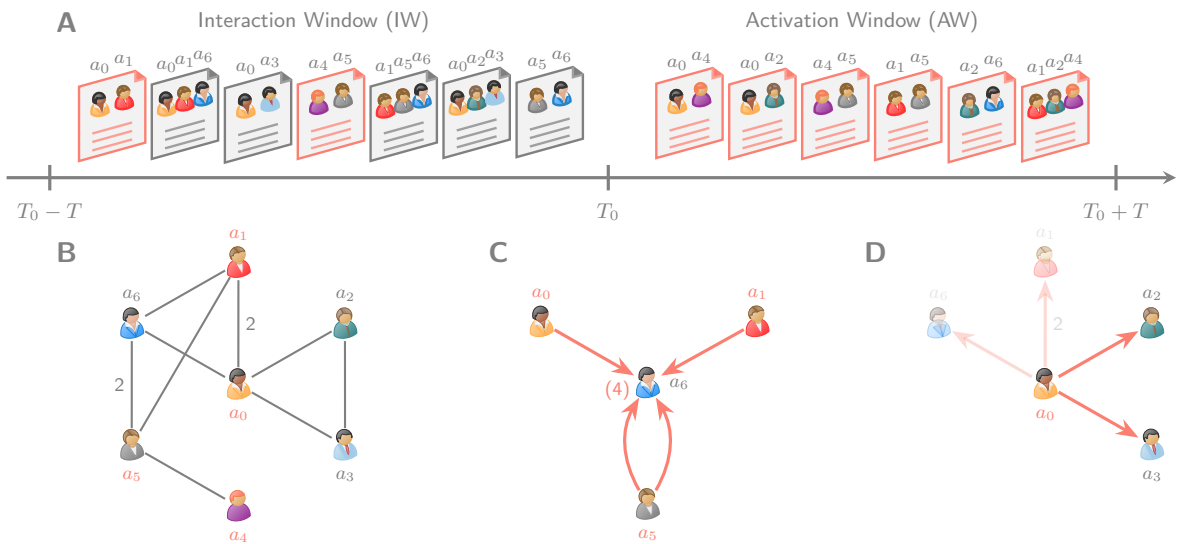

\section*{Results}
We use the scientific publication dataset OpenAlex~\citep{priem2022openalex}. We present the results for twenty topics belonging to three disciplines:  Physics, Computer Science, and Biology \& Medicine. See Methods~\ref{sec:mat-data} for details.

Our approach is inspired by the pioneering work by Kossinets and Watts on social network evolution~\cite{kossinets06}. 
In it, the authors estimated \textit{triadic closure} of two individuals $a$ and $b$, \ie, the probability that $a$ and $b$ become acquainted as a function of the number of common friends. 
They took two snapshots of the network at consecutive time ranges: in the earlier snapshot, one keeps track of all pairs of disconnected people, and in the latter, one counts how many of those pairs become connected.
A similar approach has been adopted to compute \textit{membership closure}, \ie, the probability that an individual starts participating in an activity having been connected to $k$ others who participate in it~\cite{backstrom06}. 
We now describe how we adapt this framework to measure how collaborations induce topic switches.

Given a scientific topic $t$, reference year $T_0$, and window size $T$, we construct two consecutive non-overlapping time ranges spanning years $[T_0-T, T_0)$ and $[T_0, T_0 + T)$ respectively. 
We call the first range the \textit{interaction window} (IW), where we track author interactions in the collaboration network, and the latter range, the \textit{activation window} (AW), where we count topic switches.
We then identify the set of \textit{active} authors $A$ who published papers $P$ on topic $t$ during the IW. 
For example, in Fig.~\ref{fig:toy-example}A,  $A = \{a_0, a_1, a_4, a_5\}$.
We construct the collaboration network $G$ by considering all papers $P^\prime$ written by authors $a \in A$ after $a$ becomes active. Note that $P^\prime$ includes papers outside of $P$, like the ones drawn in gray in Fig.~\ref{fig:toy-example}A.
We classify the non-active authors in $G$ as \textit{inactive} authors who are the candidates for topic switches in the AW. 
They turn active when they publish their first paper on topic $t$.
In Fig.~\ref{fig:toy-example}B, authors $a_2$, $a_3$, and $a_6$ are inactive, with $a_2$ and $a_6$ becoming active in the AW. 
Furthermore, we rank each active author $a \in A$ based on two metrics of scientific prominence: \textit{productivity} and \textit{impact}, described in the Methods~\ref{sec:ranking}, and calculated at the end of the IW to capture the current perception of $a$'s scholarly output.
Finally, for each metric, we identify and mark the authors who rank in the top and the bottom 10\%.

Given this general setup, we conduct two complementary experiments that we describe in detail in Sections \ref{sec:exp1} and \ref{sec:exp2}. 
In Experiment I, we measure membership closure among inactive authors to quantitatively assess how past collaborations with active authors manifest in topic switches. 
In Experiment II, we instead focus on the active authors, quantifying the propensity of their inactive coauthors to start working on their topic of expertise.

\begin{figure*}[htb]
    \centering
    \input{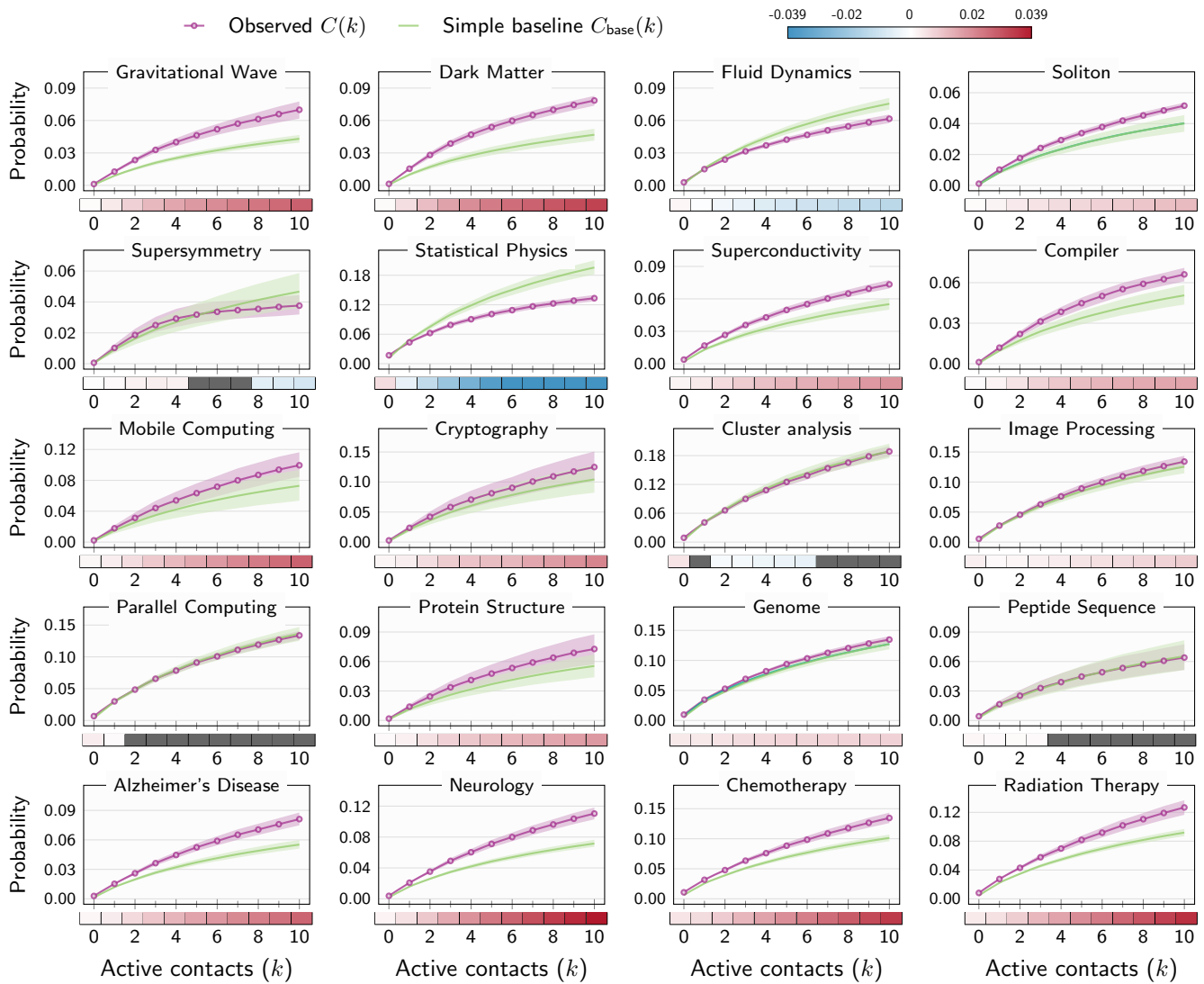}
    \caption{Experiment I. Cumulative target activation probability (in purple) for inactive authors in the AW with shaded 95\% confidence intervals. For each $k$, the $y$-value indicates the fraction of inactive authors with at least $k$ active contacts in the IW who became active in the AW.
    The green solid line with shaded errors represents the baseline described in the text, corresponding to independent effects from the coauthors.
    The heatmap below the $x$-axis shows the mean difference between the observed and baseline curves for each $k$ value. It is gray if the 95\% confidence interval contains 0, denoting the $k$-values where the points are statistically indistinguishable at $p$-value $0.05$. Positive and negative deviations from the baseline are in red and blue, respectively.}
    \label{fig:membership-closure}
\end{figure*}
 
\subsection{Experiment I} \label{sec:exp1}

Here we investigate membership closure among inactive authors. Specifically, we will answer the following questions:
\begin{itemize}
    \item How is the probability of topic switches related to $k$, the number of contacts with active authors?
    
    \item Does this probability depend on the relative prominence of the active authors?
\end{itemize}
To compute the measure, we first must define what construes as contact with an active author in the IW. We consider two definitions as described below. 

\begin{enumerate}
    \item The number of active coauthors, with the same coauthor counted as many times as the number of collaborations.
    In the collaboration network, this corresponds to the weighted degree when considering only active coauthors. 
    
    \item The number of papers written with active coauthors.     
\end{enumerate}
For example, in Fig.~\ref{fig:toy-example}C author $a_6$ has four contacts based on the first definition (two from $a_5$ and one each from $a_0$ and $a_1$), and two if we use the second (from the second and the fourth papers in the IW).
We report the findings based on the first definition in the main text. 
Results from the second definition do not alter the main conclusions and can be found in SI \Cref{fig:supp-closure,fig:supp-heatmaps}. 

To address the first question, we compute the cumulative \textit{target activation probability} $C(k)$, \ie, the fraction of inactive authors who become active in the AW as a function of the number of contacts $k$ (see Methods~\ref{sec:closure}).
In Fig.~\ref{fig:membership-closure}, we plot $C(k)$ (in purple) for each of the twenty topics under investigation. Error bars derive from averaging over different time windows for each field (see Methods~\ref{sec:stat-diff}).
As expected, we see an increasing trend. In particular, the jump from $k$ = 0 to $k$ = 1 is remarkable, showing that the probability of \textit{spontaneous} activation in the absence of previous contacts ($k$ = 0) is much lower than that of activation through collaboration ($k$ $\geq$ 1).
We observe that the higher the number of contacts, the larger the probability. 
Most of the growth occurs for low values of $k$.

To put these numbers in context, we consider a simple baseline $C_\text{base}(k)$ (see Methods~\ref{sec:baseline}) where we assume each contact has a constant, independent probability of producing a topic switch.%, independently of the others.  
Within each topic, we compute the difference (see Methods~\ref{sec:stat-diff}) between the curves for each value of $k$ over all reference years and plot them below the $x$-axis.
Except for the topics of Cluster Analysis, Parallel Computing, and Peptide Sequence, the observed curves deviate from the baseline.
This provides some empirical evidence to ascertain that the baseline cannot capture the nuances in the observed data. 
A positive deviation for the majority of the topics indicates a compounding effect.
Fluid Dynamics and Statistical Physics are exceptions, as they undershoot the baseline. This may be because they are broad interdisciplinary fields unlike the others, and having collaborators in different fields may lessen their effect.

Next, we explore the second research question, checking if the contact source's prominence affects activation chances.
Recall that in every IW for a topic, we select active authors in the top 10\% and the bottom 10\% based on productivity and impact.
This separates the most prominent active authors from the least prominent. 
To mitigate confounding effects, we only consider the subset of inactive authors who are neighbors with strictly one of the two sets of active authors. 

\begin{figure}[htb]
    \centering
    \input{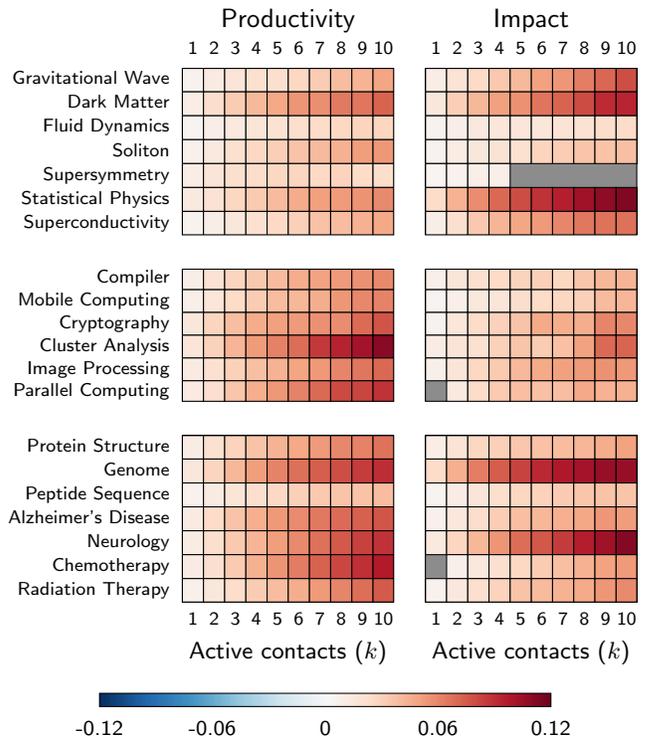}

% \pgfplotsset{colormap/PuOr}
\pgfplotsset{colormap/RdBu}
% \pgfplotsset{colormap/temp}

\begin{tikzpicture}
	\begin{groupplot}[
		group style={
			group size=2 by 3,
            group name=mygroup,
% 			vertical sep=12pt,
			vertical sep=13pt,
            horizontal sep=12pt,
			% x descriptions at=edge bottom,
            xlabels at=edge top,
            xticklabels at=edge bottom,
			y descriptions at=edge left,
		},
		width=0.245\textwidth,
        height=108pt,
		enlarge x limits={rel=0.05, upper},
		enlarge y limits={rel=0.08},
        typeset ticklabels with strut,
        title style={yshift=0.5em, align=center},
		colormap name=RdBu,
		ytick=data,
        % colorbar,
        tick align=outside,
		xtick=data,
        xtick style={draw=none},
		ytick style={draw=none, font=\scriptsize\sffamily, xshift=-5},
		xticklabel style={draw=none, font=\scriptsize\sffamily},
        yticklabel style={draw=none, font=\scriptsize\sffamily},
		axis background/.style={fill=gray!90}
	]
	%%%%%%%%% Physics
	\nextgroupplot[
		title={Productivity},
		ylabel={},
        point meta min=-0.12,
		point meta max=0.12,
        symbolic x coords={1,2,3,4,5,6,7,8,9,10},
        symbolic y coords={Gravitational wave,Dark matter,Fluid dynamics,Soliton,Supersymmetry,Statistical physics,Superconductivity},
        yticklabels={Gravitational Wave,Dark Matter,Fluid Dynamics,Soliton,Supersymmetry,Statistical Physics,Superconductivity},
        xticklabel pos=top,
        xticklabel style={yshift=-0.5em},
        yticklabel style={xshift=0.3em},
	]
	\addplot[
		matrix plot,  % not matrix plot* 
		draw=black, 
		fill opacity=1,
		colorbar source,
		 % nodes near coords,%=\coordindex,
		% mark=none,
		mesh/cols=10,
		point meta=explicit,
	] 
	table[x=k, y=topic, meta expr=-\thisrow{value}] {\PhysicsHeatmapProd};	
	
    %%%%%%%%%
	\nextgroupplot[
		title={Impact},
		ylabel={},
        point meta min=-0.12,
		point meta max=0.12,
        symbolic x coords={1,2,3,4,5,6,7,8,9,10},
		symbolic y coords={Gravitational wave,Dark matter,Fluid dynamics,Soliton,Supersymmetry,Statistical physics,Superconductivity},
        yticklabels={},	
        xticklabel pos=top,
        xticklabel style={yshift=-0.5em},
	]
	\addplot[
		matrix plot,
		draw=black, 
		fill opacity=1,
		colorbar source,
		mesh/cols=10,
		point meta=explicit,
	] 
	table[x=k, y=topic, meta expr=-\thisrow{value}] {\PhysicsHeatmapImpOne};	

    %%%%%%%%% CS
	\nextgroupplot[
		height=95pt,
        title={},
		ylabel={},
        point meta min=-0.12,
		point meta max=0.12,
        % colorbar,
		symbolic x coords={1,2,3,4,5,6,7,8,9,10},
		symbolic y coords={Compiler,Mobile computing,Cryptography,Cluster analysis,Image processing,Parallel computing},
        xticklabels={},       
        yticklabels={Compiler, Mobile Computing, Cryptography, Cluster Analysis, Image Processing, Parallel Computing},
        yticklabel style={xshift=0.3em},
	]
	\addplot[
		matrix plot,
		draw=black, 
		fill opacity=1,
		colorbar source,
		mesh/cols=10,
		point meta=explicit,
	] 
	table[x=k, y=topic, meta expr=-\thisrow{value}] {\CSHeatmapProd};	
	
    %%%%%%%%%
	\nextgroupplot[
        height=95pt,
		title={},
		ylabel={},
        point meta min=-0.12,
		point meta max=0.12,
        % colorbar,
		symbolic x coords={1,2,3,4,5,6,7,8,9,10},
		symbolic y coords={Compiler,Mobile computing,Cryptography,Cluster analysis,Image processing,Parallel computing},
		% xticklabel pos=top,
        xticklabels={},
        yticklabels={},
	]
	\addplot[
		matrix plot,
		draw=black, 
		fill opacity=1,
		colorbar source,
		mesh/cols=10,
		point meta=explicit,
	] 
	table[x=k, y=topic, meta expr=-\thisrow{value}] {\CSHeatmapImpOne};

 %%%%%%%%% BioMed
	\nextgroupplot[
		title={},
		ylabel={},
        point meta min=-0.12,
		point meta max=0.12,
        % colorbar,
		symbolic x coords={1,2,3,4,5,6,7,8,9,10},
		symbolic y coords={Protein structure,Genome,Peptide sequence,Alzheimer's disease,Neurology,Chemotherapy,Radiation therapy},
        yticklabels={Protein Structure, Genome, Peptide Sequence, Alzheimer's Disease, Neurology, Chemotherapy, Radiation Therapy},
        xticklabel style={yshift=0.5em},
        yticklabel style={xshift=0.3em},
		% xticklabel pos=top,
        % xticklabels={},       
	]
	\addplot[
		matrix plot,
		draw=black, 
		fill opacity=1,
		colorbar source,
		mesh/cols=10,
		point meta=explicit,
	] 
	table[x=k, y=topic, meta expr=-\thisrow{value}] {\BioMedHeatmapProd};	
	
    %%%%%%%%%
	\nextgroupplot[
		title={},
		ylabel={},
        point meta min=-0.12,
		point meta max=0.12,
        % colorbar,
		symbolic x coords={1,2,3,4,5,6,7,8,9,10},
		symbolic y coords={Protein structure,Genome,Peptide sequence,Alzheimer's disease,Neurology,Chemotherapy,Radiation therapy},
		% xticklabel pos=top,
        % xticklabels={},
        yticklabels={},
        xticklabel style={yshift=0.5em},
	]
	\addplot[
		matrix plot,
		draw=black, 
		fill opacity=1,
		colorbar source,
		mesh/cols=10,
		point meta=explicit,
	] 
	table[x=k, y=topic, meta expr=-\thisrow{value}] {\BioMedHeatmapImpOne};

	\end{groupplot}

    \node[anchor=south, align=center] at ($(mygroup c1r3.south)+(0em, -2.9em)$) {Active contacts ($k$)};

    \node[anchor=south, align=center] at ($(mygroup c2r3.south)+(0em, -2.9em)$) {Active contacts ($k$)};

\begin{scope}[shift={(-1.1, -5.8)}, scale=1]
    \pgfplotscolorbardrawstandalone[
        colorbar horizontal,
        point meta min=-0.12, point meta max=0.12,
        colormap name=RdBu,
        colorbar style={
            x dir=reverse,
            height=5pt,
	    	xtick={-0.12,-0.06,0,0.06,0.12}, 
            xticklabels={0.12,0.06,0,-0.06,-0.12},  % need to manually 
            scale=1,
            xtick style={
                draw=none,
                /pgf/number format/.cd,
                fixed,
                fixed zerofill,
                precision=0,
                /tikz/.cd
        	},
            xticklabel style={font=\footnotesize\sffamily},
            xticklabel pos=lower,
            scaled ticks=false,
        },
    ]
\end{scope}
\end{tikzpicture}
    \caption{Heatmaps showing the mean difference between the cumulative target activation probabilities of the inactive authors in the AW who had exclusive contacts with the top 10\% and bottom 10\% of active authors, respectively, selected according to productivity (left) and impact (right) in the IW. The cells are gray if the 95\% confidence interval contains 0. The majority of red cells indicate that the cumulative target activation probabilities for contacts with the top 10\% are higher than those with the bottom 10\%.}
    \label{fig:closure-heatmaps}
\end{figure}

In Fig.~\ref{fig:closure-heatmaps}, we assess the significance of the difference between the cumulative target activation probabilities for inactive authors in contact with active authors in the two bins. 
Each row corresponds to a topic, and the color of each square indicates whether the difference is positive (red), negative (blue), or non-significant (grey). 
The two columns correspond to prominent authors selected based on productivity (left) and impact (right). For productivity, all differences are significant and positive, meaning that contacts with highly productive active authors lead to higher target activation probabilities. For impact, there are a handful of exceptions. Overall, having prominent contacts increases the target activation probability.

\subsection{Experiment II} \label{sec:exp2}

Here we focus on the active authors and their collaborators. 
For every active author $a$, we consider the subset of their inactive coauthors who have \textit{exclusively} collaborated with $a$ in the IW. 
We call this set the exclusive inactive coauthors of $a$. 
For example, in \Cref{fig:toy-example}D, active author $a_0$ has four coauthors $\{a_1, a_2, a_3, a_6\}$, of whom only $a_2$ and $a_3$ exclusively collaborate with $a_0$ in the IW. 
We do this because effects due to active authors different from $a$ would be difficult to disentangle and could confound the analysis and the conclusions. 
The relevant measure here is the \textit{source activation probability} $P_s^a$, \ie, the fraction of exclusive inactive coauthors who become active in the AW (see Methods~\ref{sec:source-active}). 
The fraction controls for the collaboration neighborhood sizes which could vary widely for different scholars. 
In \Cref{fig:toy-example}D, $P_s^a$ for $a_0$ is   $\tfrac{1}{2}$ = 0.5, as only $a_2$ becomes active in the AW. 

For a given set of active authors, we obtain $C_s$, the complementary cumulative probability distribution of their source activation probabilities (see Methods~\ref{sec:source-active}).
We select the pools of the most and least prominent authors as described in Experiment I. 
The relative effects of the two groups are estimated by comparing the \textit{cumulative source activations}, \ie, points on the respective cumulative distributions at a specific threshold $f^\ast$. 
Results are reported in Fig.~\ref{fig:EXP2_10}A for a threshold $f^\ast = 0.10$. 
Our conclusions also hold when considering a threshold $f^\ast = 0.20$, which can be found in SI \Cref{fig:supp-exp2ab-20}.

\begin{figure}[ht!]
    \centering
    \input{data/influence}

\begin{tikzpicture}

\begin{scope}[shift={(3, 4.1)}]
    \begin{customlegend}[ 
    legend columns=4,
    legend style={
    draw=none,
    column sep=2ex,
    % nodes={scale=1.8, transform shape},
  },
  legend cell align={left},
  legend entries={Productivity~~,Impact}
  ]
    \addlegendimage{Accent-A, mark options={draw=Accent-A!70!black, thick, fill=Accent-A!70!white,}, draw opacity=0.8, only marks,}
    \addlegendimage{PiYG-E, mark options={draw=PiYG-E!70!black, thick, fill=PiYG-E!70!white}, draw opacity=0.8, only marks,}
    \end{customlegend}
\end{scope}

\begin{groupplot}[
        group style={
			group name=my plots,
			% group size=3 by 3,
            rows=3, columns=2,
			% x descriptions at=edge bottom,
            xticklabels at=all,
            xlabels at=edge bottom,
			y descriptions at=edge left,
			vertical sep=25pt,
            horizontal sep=20pt,
		},
        height=150pt,
        width=122pt,
        xmajorgrids,
        % ymajorgrids,
        major y grid style={draw=gray, draw opacity=0.4},
        ytick style={yshift=0pt},
		grid style={draw=gray!50, draw opacity=0.5},
        % xtick align=center, 
        % xtick style={draw=none},
		ytick style={draw=none},
		yticklabel style={yshift=-5pt, xshift=-12pt, font=\scriptsize\sffamily},
        axis y line*=middle,
        axis x line*=bottom, 
        enlarge y limits={0.05, upper},
        enlarge y limits={0.15, lower},
        enlarge x limits=0.05,
        title style={yshift=0em, align=center},
        xlabel={Difference}, ylabel={}, 
        axis background/.style={fill=gray!3},
    ]

%%%% Physics - Exp A
    
    \nextgroupplot[
        title={A}, 
        symbolic y coords={Superconductivity,Statistical physics,Supersymmetry,Soliton,Fluid dynamics,Dark matter,Gravitational wave},
        ytick={Superconductivity,Statistical physics,Supersymmetry,Soliton,Fluid dynamics,Dark matter,Gravitational wave},
        yticklabels={Superconductivity, Statistical Physics, Supersymmetry, Soliton, Fluid Dynamics, Dark Matter, Gravitational Wave},
        xmin=-0.02, xmax=0.1, 
        xtick={0,0.05,0.1}, xticklabels={0,0.05,0.10},
        minor xtick={0.025,0.075},
    ]
    
        \addplot [error1, error bars/.cd, x dir=both, x explicit, error bar style=thick, error mark=|, error mark options={thick},] table[x=Prod_diffA_mean, y=topic, x error plus expr=\thisrow{Prod_diffA_mean_ci95_max}-\thisrow{Prod_diffA_mean}, x error minus expr=\thisrow{Prod_diffA_mean}-\thisrow{Prod_diffA_mean_ci95_min}] {\PhysicsInfluenceProdImpOneTen}; 
        %\addlegendentry{Productivity};

        \addplot [error2, error bars/.cd, x dir=both, x explicit, error bar style=thick, error mark=|, error mark options={thick},] table[x=Imp1_diffA_mean, y=topic, x error plus expr=\thisrow{Imp1_diffA_mean_ci95_max}-\thisrow{Imp1_diffA_mean}, x error minus expr=\thisrow{Imp1_diffA_mean}-\thisrow{Imp1_diffA_mean_ci95_min}] {\PhysicsInfluenceProdImpOneTen}; 
        %\addlegendentry{Impact};

        % \addplot[mark=none, draw opacity=0.5, draw=gray] coordinates {(0,current bounding box.south) (0,Statistical physics)};
        % \draw[add=2 and .5] (axis cs:0,Gravitational wave) -- (axis cs:0,Statistical physics);

    %%%%% Exp B

    \nextgroupplot[
        title={B}, 
        symbolic y coords={Superconductivity,Statistical physics,Supersymmetry,Soliton,Fluid dynamics,Dark matter,Gravitational wave}, 
        ytick={Superconductivity,Statistical physics,Supersymmetry,Soliton,Fluid dynamics,Dark matter,Gravitational wave},
        yticklabels=\empty,
        xmin=-0.05, xmax=0.3, 
        xtick={0,0.1,0.2,0.3}, xticklabels={0,0.10,0.20,0.30},
        minor xtick={0.05,0.15,0.25},
    ]
        \addplot [error1, error bars/.cd, x dir=both, x explicit, error bar style=thick, error mark=|, error mark options={thick},] table[x=Prod_diffB_mean, y=topic, x error plus expr=\thisrow{Prod_diffB_mean_ci95_max}-\thisrow{Prod_diffB_mean}, x error minus expr=\thisrow{Prod_diffB_mean}-\thisrow{Prod_diffB_mean_ci95_min}] {\PhysicsInfluenceProdImpOneTen};         

        \addplot [error2, error bars/.cd, x dir=both, x explicit, error bar style=thick, error mark=|, error mark options={thick},] table[x=Imp1_diffB_mean, y=topic, x error plus expr=\thisrow{Imp1_diffB_mean_ci95_max}-\thisrow{Imp1_diffB_mean}, x error minus expr=\thisrow{Imp1_diffB_mean}-\thisrow{Imp1_diffB_mean_ci95_min}] {\PhysicsInfluenceProdImpOneTen};

    %%%%%  CS Exp A

    \nextgroupplot[
        title={},
        symbolic y coords={Parallel computing,Image processing,Cluster analysis,Cryptography,Mobile computing,Compiler},
        ytick={Parallel computing,Image processing,Cluster analysis,Cryptography,Mobile computing,Compiler},
        yticklabels={Parallel Computing,Image Processing,Cluster Analysis,Cryptography,Mobile Computing,Compiler},
        xmin=-0.02, xmax=0.1, 
        xtick={0,0.05,0.10}, xticklabels={0,0.05,0.10},
        minor xtick={0.025,0.075},
    ]

        \addplot [error1, error bars/.cd, x dir=both, x explicit, error bar style=thick, error mark=|, error mark options={thick},] table[x=Prod_diffA_mean, y=topic, x error plus expr=\thisrow{Prod_diffA_mean_ci95_max}-\thisrow{Prod_diffA_mean}, x error minus expr=\thisrow{Prod_diffA_mean}-\thisrow{Prod_diffA_mean_ci95_min}] {\CSInfluenceProdImpOneTen}; 

        \addplot [error2, error bars/.cd, x dir=both, x explicit, error bar style=thick, error mark=|, error mark options={thick},] table[x=Imp1_diffA_mean, y=topic, x error plus expr=\thisrow{Imp1_diffA_mean_ci95_max}-\thisrow{Imp1_diffA_mean}, x error minus expr=\thisrow{Imp1_diffA_mean}-\thisrow{Imp1_diffA_mean_ci95_min}] {\CSInfluenceProdImpOneTen};
        
    %%%%%  CS Exp B

    \nextgroupplot[
        title={},
        symbolic y coords={Parallel computing,Image processing,Cluster analysis,Cryptography,Mobile computing,Compiler}, 
        ytick={Parallel computing,Image processing,Cluster analysis,Cryptography,Mobile computing,Compiler},
        yticklabels=\empty,
        xmin=-0.05, xmax=0.3, 
        xtick={0,0.10,0.20,0.30}, xticklabels={0,0.10,0.20,0.30},
        minor xtick={0.05,0.15,0.25},
    ]
        
        \addplot [error1, error bars/.cd, x dir=both, x explicit, error bar style=thick, error mark=|, error mark options={thick},] table[x=Prod_diffB_mean, y=topic, x error plus expr=\thisrow{Prod_diffB_mean_ci95_max}-\thisrow{Prod_diffB_mean}, x error minus expr=\thisrow{Prod_diffB_mean}-\thisrow{Prod_diffB_mean_ci95_min}] {\CSInfluenceProdImpOneTen};
        
        \addplot [error2, error bars/.cd, x dir=both, x explicit, error bar style=thick, error mark=|, error mark options={thick},] table[x=Imp1_diffB_mean, y=topic, x error plus expr=\thisrow{Imp1_diffB_mean_ci95_max}-\thisrow{Imp1_diffB_mean}, x error minus expr=\thisrow{Imp1_diffB_mean}-\thisrow{Imp1_diffB_mean_ci95_min}] {\CSInfluenceProdImpOneTen};

    %%%% BioMed Exp A
    
    \nextgroupplot[
        title={}, 
        symbolic y coords={Radiation therapy,Chemotherapy,Neurology,Alzheimer's disease,Peptide sequence,Genome,Protein structure}, 
        ytick={Radiation therapy,Chemotherapy,Neurology,Alzheimer's disease,Peptide sequence,Genome,Protein structure},
        yticklabels={Radiation Therapy,Chemotherapy,Neurology,Alzheimer's Disease,Peptide Sequence,Genome,Protein Structure}, 
        xmin=-0.02, xmax=0.1, 
        xtick={0,0.05,0.1}, xticklabels={0,0.05,0.10},
        minor xtick={0.025,0.075},
    ]
        
        \addplot [error1, error bars/.cd, x dir=both, x explicit, error bar style=thick, error mark=|, error mark options={thick},] table[x=Prod_diffA_mean, y=topic, x error plus expr=\thisrow{Prod_diffA_mean_ci95_max}-\thisrow{Prod_diffA_mean}, x error minus expr=\thisrow{Prod_diffA_mean}-\thisrow{Prod_diffA_mean_ci95_min}] {\BioMedInfluenceProdImpOneTen}; 

        \addplot [error2, error bars/.cd, x dir=both, x explicit, error bar style=thick, error mark=|, error mark options={thick},] table[x=Imp1_diffA_mean, y=topic, x error plus expr=\thisrow{Imp1_diffA_mean_ci95_max}-\thisrow{Imp1_diffA_mean}, x error minus expr=\thisrow{Imp1_diffA_mean}-\thisrow{Imp1_diffA_mean_ci95_min}] {\BioMedInfluenceProdImpOneTen}; 

    %%%%%%%%%% Exp B
    \nextgroupplot[
        title={},
        symbolic y coords={Radiation therapy,Chemotherapy,Neurology,Alzheimer's disease,Peptide sequence,Genome,Protein structure}, 
        ytick={Radiation therapy,Chemotherapy,Neurology,Alzheimer's disease,Peptide sequence,Genome,Protein structure},
        yticklabels=\empty, 
        xmin=-0.05, xmax=0.3, 
        xtick={0,0.1,0.2,0.3}, xticklabels={0,0.10,0.20,0.30},
        minor xtick={0.05,0.15,0.25},
    ]

        \addplot [error1, error bars/.cd, x dir=both, x explicit, error bar style=thick, error mark=|, error mark options={thick},] table[x=Prod_diffB_mean, y=topic, x error plus expr=\thisrow{Prod_diffB_mean_ci95_max}-\thisrow{Prod_diffB_mean}, x error minus expr=\thisrow{Prod_diffB_mean}-\thisrow{Prod_diffB_mean_ci95_min}] {\BioMedInfluenceProdImpOneTen};

        \addplot [error2, error bars/.cd, x dir=both, x explicit, error bar style=thick, error mark=|, error mark options={thick},] table[x=Imp1_diffB_mean, y=topic, x error plus expr=\thisrow{Imp1_diffB_mean_ci95_max}-\thisrow{Imp1_diffB_mean}, x error minus expr=\thisrow{Imp1_diffB_mean}-\thisrow{Imp1_diffB_mean_ci95_min}] {\BioMedInfluenceProdImpOneTen};

\end{groupplot}
\end{tikzpicture}
    \caption{
    Experiment II results for $f^\ast = 0.10$. (A) The mean and 95\% confidence interval of the means of the difference between the cumulative source activations of active authors in the top 10\% and bottom 10\% based on productivity (green) and impact (pink). (B) The mean and 95\% confidence interval of the means of the difference between the chaperoning propensities of active authors in the top 10\% and bottom 10\% based on productivity (green) and impact (pink). A positive difference indicates that the effect is stronger for the top 10\% active authors.
    % Combined difference plot for Exp 2A, 2B for threshold \textbf{0.10}. The two sets are indistinguishable at $p=0.05$ if the confidence intervals contain $0$.
    }
    \label{fig:EXP2_10}
\end{figure}

  % to remove the single line in the previous page
In \Cref{fig:EXP2_10}, each row corresponds to a topic. The green and purple ranges represent the $95\%$ confidence intervals of the mean difference between the cumulative source activations for the two pools of authors for productivity and impact, respectively. For productivity, the difference is significant for all topics but one (Superconductivity). 
The differences are somewhat less pronounced for impact, but are still significant in most cases.

To further corroborate this finding, we specialize the analysis by checking how many exclusive coauthors of $a$ also published their first paper on topic $t$ in the AW with $a$. 
This is a way to assess the \textit{chaperoning propensity} of active authors~\citep{sekara2018chaperone}, and we define the measure in  Methods~\ref{sec:chaperone}.  
In Fig.~\ref{fig:EXP2_10}B, we report the $95\%$ confidence intervals of the average difference between the chaperoning propensities for the most prominent and the least prominent active authors for threshold $f^\ast = 0.10$.
Similar to \Cref{fig:EXP2_10}A, we find that the more productive/impactful an active author is, the more likely their coauthors will start working with them on a new topic.
Results for $f^\ast = 0.20$, which confirm this trend, can be found in SI \Cref{fig:supp-exp2ab-20}.

While our analysis clearly shows that prominence is a factor, one may wonder if the number of coauthors also plays a role. 
We posit that, on average, the more collaborators one has, the more tenuous the contact with any of them will be, resulting in lower source activation probabilities. 
% Therefore, we check whether the number of collaborators also affects the ability of a prominent active author to activate their inactive coauthors. 
From each group of most prominent authors, we, therefore, pick the top and the bottom 20\% based on the average number of coauthors on papers published with exclusive inactive coauthors. 
By construction, this excludes any paper written on the focal topic.

\begin{figure*}[htb!]
    \centering
    \input{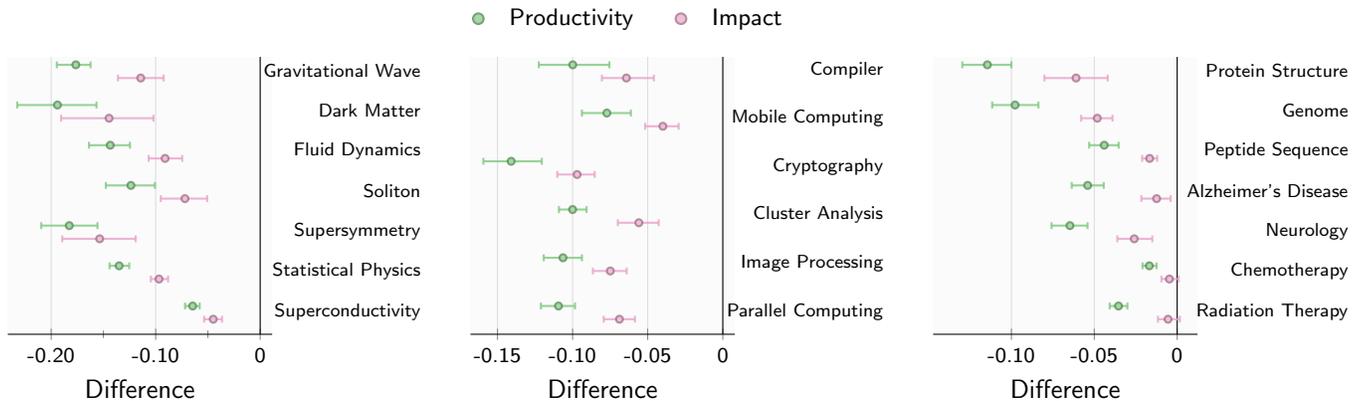}

\begin{tikzpicture}

\begin{scope}[shift={(9.5, 3.5)}]
    \begin{customlegend}[ 
    legend columns=2,
    legend style={
    draw=none,
    column sep=2ex,
    % nodes={scale=1.8, transform shape},
  },
  legend cell align={left},
  legend entries={Productivity~~~,Impact}
  ]
    \addlegendimage{Accent-A, mark options={draw=Accent-A!70!black, thick, fill=Accent-A!70!white,}, draw opacity=0.8, only marks,}
    \addlegendimage{PiYG-E, mark options={draw=PiYG-E!70!black, thick, fill=PiYG-E!70!white}, draw opacity=0.8, only marks,}
    \end{customlegend}
\end{scope}

\begin{groupplot}[
        group style={
			group name=my plots,
			% group size=3 by 3,
            rows=1, columns=3,
			% x descriptions at=edge bottom,
            xticklabels at=all,
            xlabels at=edge bottom,
			y descriptions at=edge left,
			vertical sep=25pt,
            % horizontal sep=15pt,
            horizontal sep=75pt,
		},
        height=150pt,
        width=145pt,
        xmajorgrids,
        % ymajorgrids,
        major y grid style={draw=gray, draw opacity=0.4},
        ytick style={yshift=0pt},
		grid style={draw=gray!50, draw opacity=0.5},
        % xtick align=center, 
        % xtick style={draw=none},
		ytick style={draw=none},
		yticklabel style={yshift=-5pt, xshift=68pt, font=\scriptsize\sffamily},
        % axis y line*=none,
        axis x line*=bottom,
        axis y line*=middle,
        % every inner x axis line/.append style={--},
        % y axis line style={draw=none},
        % enlarge y limits={0.01, upper},
        enlarge y limits={0.05, upper},
        enlarge y limits={0.15, lower},
        enlarge x limits=0.05, 
        title style={yshift=0em, align=center},
        xlabel={Difference}, ylabel={}, 
        axis background/.style={fill=gray!3},
    ]

    %%%% Physics Exp C 
    
    \nextgroupplot[
        title={}, 
        yticklabel style={xshift=-4pt},
        symbolic y coords={Superconductivity,Statistical physics,Supersymmetry,Soliton,Fluid dynamics,Dark matter,Gravitational wave}, 
        ytick={Superconductivity,Statistical physics,Supersymmetry,Soliton,Fluid dynamics,Dark matter,Gravitational wave},
        yticklabels={Superconductivity, Statistical Physics, Supersymmetry, Soliton, Fluid Dynamics, Dark Matter, Gravitational Wave}, 
        xmin=-0.23, xmax=0, 
        minor x tick num=1,
        xtick={-0.2,-0.1,0}, xticklabels={-0.20,-0.10,0},
        % minor xtick={0.05,0.15,0.25},
    ]
    
        \addplot [error1, error bars/.cd, x dir=both, x explicit, error bar style=thick, error mark=|, error mark options={thick},] table[x=Prod_diffC_mean, y=topic, x error plus expr=\thisrow{Prod_diffC_mean_ci95_max}-\thisrow{Prod_diffC_mean}, x error minus expr=\thisrow{Prod_diffC_mean}-\thisrow{Prod_diffC_mean_ci95_min}] {\PhysicsInfluenceProdImpOneTen};

        \addplot [error2, error bars/.cd, x dir=both, x explicit, error bar style=thick, error mark=|, error mark options={thick},] table[x=Imp1_diffC_mean, y=topic, x error plus expr=\thisrow{Imp1_diffC_mean_ci95_max}-\thisrow{Imp1_diffC_mean}, x error minus expr=\thisrow{Imp1_diffC_mean}-\thisrow{Imp1_diffC_mean_ci95_min}] {\PhysicsInfluenceProdImpOneTen};

    %%% CS Exp C
    \nextgroupplot[
        title={}, 
        yticklabel style={xshift=-4pt},
        symbolic y coords={Parallel computing,Image processing,Cluster analysis,Cryptography,Mobile computing,Compiler}, 
        ytick={Parallel computing,Image processing,Cluster analysis,Cryptography,Mobile computing,Compiler}, 
        yticklabels={Parallel Computing, Image Processing, Cluster Analysis, Cryptography, Mobile Computing, Compiler},
        xmin=-0.16, xmax=0, 
        % minor x tick num=1,
        xtick={-0.15,-0.1,-0.05,0}, xticklabels={-0.15,-0.10,-0.05,0},
        % minor xtick={0.05,0.15,0.25},
    ]

        \addplot [error1, error bars/.cd, x dir=both, x explicit, error bar style=thick, error mark=|, error mark options={thick},] table[x=Prod_diffC_mean, y=topic, x error plus expr=\thisrow{Prod_diffC_mean_ci95_max}-\thisrow{Prod_diffC_mean}, x error minus expr=\thisrow{Prod_diffC_mean}-\thisrow{Prod_diffC_mean_ci95_min}] {\CSInfluenceProdImpOneTen};

        \addplot [error2, error bars/.cd, x dir=both, x explicit, error bar style=thick, error mark=|, error mark options={thick},] table[x=Imp1_diffC_mean, y=topic, x error plus expr=\thisrow{Imp1_diffC_mean_ci95_max}-\thisrow{Imp1_diffC_mean}, x error minus expr=\thisrow{Imp1_diffC_mean}-\thisrow{Imp1_diffC_mean_ci95_min}] {\CSInfluenceProdImpOneTen};

    % %%%%%%% BioMed Exp C
    \nextgroupplot[
        title={},
        symbolic y coords={Radiation therapy,Chemotherapy,Neurology,Alzheimer's disease,Peptide sequence,Genome,Protein structure},
        ytick={Radiation therapy,Chemotherapy,Neurology,Alzheimer's disease,Peptide sequence,Genome,Protein structure},
        yticklabels={Radiation Therapy, Chemotherapy, Neurology, Alzheimer's Disease, Peptide Sequence, Genome, Protein Structure},
        xmin=-0.14, xmax=0.005, 
        % minor x tick num=1,
        xtick={-0.15,-0.1,-0.05,0}, xticklabels={-0.15,-0.10,-0.05,0},
        % minor xtick={0.05,0.15,0.25},
    ]

        \addplot [error1, error bars/.cd, x dir=both, x explicit, error bar style=thick, error mark=|, error mark options={thick},] table[x=Prod_diffC_mean, y=topic, x error plus expr=\thisrow{Prod_diffC_mean_ci95_max}-\thisrow{Prod_diffC_mean}, x error minus expr=\thisrow{Prod_diffC_mean}-\thisrow{Prod_diffC_mean_ci95_min}] {\BioMedInfluenceProdImpOneTen};

        \addplot [error2, error bars/.cd, x dir=both, x explicit, error bar style=thick, error mark=|, error mark options={thick},] table[x=Imp1_diffC_mean, y=topic, x error plus expr=\thisrow{Imp1_diffC_mean_ci95_max}-\thisrow{Imp1_diffC_mean}, x error minus expr=\thisrow{Imp1_diffC_mean}-\thisrow{Imp1_diffC_mean_ci95_min}] {\BioMedInfluenceProdImpOneTen};

\end{groupplot}
\end{tikzpicture}
    \caption{Dilution effect results for $f^\ast = 0.10$. The mean and 95\% confidence interval of the mean of the difference between the cumulative source activations of active authors in the top 20\% and bottom 20\% bins, based on the average number of coauthors, among the top 10\% active authors in productivity (green) and impact (pink).
    A negative difference across the topics indicates a \textit{dilution} effect, wherein coauthors of prominent active scholars with fewer collaborators are more likely to switch topics.}
    \label{fig:EXP_2C}
\end{figure*}

In Fig.~\ref{fig:EXP_2C}, we perform the same analysis as in Fig.~\ref{fig:EXP2_10}A for the two pools of authors described above. 
We observe that the confidence intervals of the differences lie to the left of zero, \ie, are negative. 
For productivity, all values are significant. For impact, there are only two topics (Chemotherapy and Radiation Therapy) that are not significant. 
Overall, inactive coauthors of prominent authors with more collaborators have a lower probability of switching topics. 
This is consistent with the intuition that the interactions with each coauthor are less frequent/strong in that case and, consequently, less effective at inducing topic switches.

\section*{Discussion}

Collaboration allows scholars to deepen existing knowledge and be exposed to new ideas. %, which could lead to topic switches.
In this paper, we assessed if and how collaboration patterns affect the probability of switching research topics. 
We determined that the probability for a scholar to start working on a new topic depends on earlier contacts with people already active in that topic. 
This effect is proportional to the number of contacts, with more contacts resulting in higher probabilities. 
In most topics, this behavior is distinct from a simple baseline assuming independent effects from the contacts, which likely indicates effects of non-dyadic interactions that prompt further investigation.

Similarly, we measured the probability that inactive coauthors of an active author end up publishing on the new topic, which singles out the effect of the association with that author in the activation process. 
We stress that, by design, previous interactions between inactive and active authors are limited to works dealing with topics different from the focal topic. Therefore, our analysis suggests that an active author may expose an inactive one to a new topic, even when their interactions do not directly concern that topic. This underlines the social character of scientific interactions, where discussions may deviate from the context that mainly motivates them. 

We also checked whether the activation probability depends on some specific features of the active authors. We found that the more prolific and impactful authors have higher chances of inducing coauthors to switch topics and become coauthors in their first paper on the new topic.

Furthermore, we showed that the larger the number of coauthors of an active author, the lower the chance of a topic switch. This is consistent with a \textit{dilution} of the influence, resulting from the inability to interact strongly with collaborators when their number is large. To the best of our knowledge, we are disclosing this effect for the first time.

A natural explanation of our findings is that topic switches result from a social contagion process, much like the adoption of new products~\cite{bass69,leskovec07}, or the spreading of political propaganda~\cite{bond12}. 
However, we cannot discount selection effects in observational studies like ours~\citep{shalizi11}. Having large numbers of active coauthors on a topic may be associated with  strong latent homophily between the authors, which may facilitate the future adoption of the topic even without interventions from the active authors. 

Our work uses OpenAlex, a valuable open-access bibliometric database. 
We rely on their author disambiguation and topic classification algorithms to conduct the analyses. 
These processes are inherently noisy and can introduce implicit biases. 
In addition, there appears to be incomplete citation coverage which might partly explain why the results for impact are less robust than those for productivity.
Future releases of OpenAlex might mitigate these problems.
To counter these issues, we repeated our analysis on multiple topics from three distinct scientific disciplines. 
While the size of the effects varies with the topic, our main conclusions hold across all topics, with very few exceptions. 

In conclusion, our work offers a platform for further investigations on the mechanisms driving homophily in science. A thorough understanding of these mechanisms requires effective integration of all factors that may play a role.
Besides productivity and impact, topic switches may be affected by the institutional affiliations of those involved. On the one hand, it is plausible that people in the same institution have more chances to interact and affect each other's behavior. On the other hand, collaborations with people from renowned institutions are expected to weigh more in the process.
Another discriminating factor could be the number of citations to the collaborator's papers. The higher the number of citations, the closer the association between collaborators.
We could also include the scientific affinity between coauthors through the similarity of their papers. Modern neural language models~\cite{mikolov13,devlin18} allow to embed papers and, consequently, authors in high-dimensional vector spaces, where the distance between two authors is a good proxy of the similarity of their outputs.

\section*{Methods}
\setcounter{subsection}{0}  % for resetting subsection counter for Methods

\subsection{Data} \label{sec:mat-data}
We analyze papers from the February 2023 snapshot of the bibliometric dataset OpenAlex: the successor to Microsoft Academic Graph (MAG). We restrict our analysis to papers published between 1990 and 2022 and having at most thirty authors. Papers are tagged with \textit{concepts} (topics) by a classifier trained on the MAG. %Concepts represent the topics of the papers. %There are about 65,000 total concepts laid out as a tree across six levels. 
We use concept tags to construct snapshots for three fields: Physics, Computer Science (CS), and Biology and Medicine (BioMed). 
Physics contains 19.7M papers, while CS and BioMed each have 27.6M and 43.52M papers, respectively. 
Within each domain, we select seven, six, and seven topics, respectively. 
We publish the code and associated data on \href{https://github.com/satyakisikdar/Homophily-Topic-Switches}{GitHub}.

Within each topic, we consider reference years between 1995 and 2018, where the respective interaction and activation windows contain at least 3000 papers.
This threshold ensures a critical mass of papers and authors to conduct the analyses. 
Each topic we selected has at least 10 reference years satisfying the constraint. 
The statistical tests in the manuscript are aggregated over the different reference years. 
More information is available in SI~\Cref{tab:Physics-info,tab:CS-info,tab:BioMed-info}.
%The errors on the estimates of the activation probabilities are standard errors of the mean, obtained by averaging over all time windows for each given field.

\subsection{Overlap coefficient}
\label{sec-overl}

We use the overlap coefficient to measure the degree of overlap between the different sets of authors picked based on productivity and impact.

$$
    \text{Overlap}(A, B) = \frac{|A \cap B|}{\min(|A|, |B|)}.
$$

In our case, the two sets are the same size, so a score of 10\% implies that both sets share 10\% of the elements. 

\subsection{Author ranking metrics} \label{sec:ranking}
Let $P$ be the set of papers published on topic $t$ authored by the set of active authors $A$ during the interaction window IW. 
Let $a$ be an active author who wrote $P_a$ papers during the IW. 
We define the following metrics to rank active authors and select the top and bottom 10\%.

\textit{Productivity:} the count of papers $a$ has authored on topic $t$ during the IW. More formally, it is the cardinality of the set $P \cap P_a$. 

\textit{Impact:} the average citation count of $P_a$ from the papers in $P$. 
We argue that restricting incoming citations from $P$ is a good proxy for the impact that $a$ has made on that topic. The average number of citations is a better indicator of excellence than the total citation count~\citep{erkol23}. Also,
considering the average instead of the sum lowers its correlation with productivity, here measured by the overlap coefficient of Methods~\ref{sec-overl}, as often the most productive authors are also the most cited ones~\citep{zeng22}. 
A low correlation lets us safely disregard the confounding effects of the two metrics and allows us to treat them as fairly independent variables.
Correlation statistics are reported in SI~\Cref{tab:Physics-overlap,tab:CS-overlap,tab:BioMed-overlap}.

\subsection{Statistical test for difference of samples} \label{sec:stat-diff}
To test whether two independent samples $X_1$ and $X_2$ are different concerning their means $\mu_1$ and $\mu_2$, we assume the null hypothesis $H_0: \mu_1 = \mu_2$.
We compute the mean and 95\% confidence interval of $\mu_1 - \mu_2$ using bootstrapping and reject the null hypothesis $H_0$ at $p < 0.05$ if the confidence interval \textit{does not} contain $0$~\citep{gardner1986confidence}.
In other words, $X_1$ and $X_2$ are considered statistically different at $p < 0.05$ if the 95\% confidence interval of the difference of their respective means does not contain $0$. 
Furthermore, a positive mean of the difference indicates that $X_1 > X_2$, while a negative mean indicates $X_1 < X_2$.

\subsection{Target activation probability} \label{sec:closure}
% In Experiment I, we investigate if having multiple contacts with active scholars makes an inactive author more likely to become active during the observation window. 
Let $n(k)$ be the number of inactive authors with exactly $k$ contacts during the exposure window, of whom $m(k)$ become active in the observation window. 
The \textit{target activation probability} $P(k)$ is the probability of becoming active after having exactly $k$ contacts, defined as
\begin{equation}
    P(k) = \frac{m(k)}{n(k)}.
    \label{eq:watts-actual}
\end{equation} 

% We aggregate authors' probabilities in cumulative bins to increase their statistical significance.
The \textit{cumulative target activation probability} $C(k)$ with $k$ or more contacts is given by
\begin{equation}
    C(k) = \tfrac{\sum_k^\infty m(k)}{\sum_k^\infty n(k)}.   
    \label{eq:watts-cumulative}
\end{equation}

\subsection{Simple baseline for membership closure} \label{sec:baseline}
Let $p$ represent the probability of activation from a single contact. The probability of activation having $k$ contacts, acting independently of each other, is $P_\text{base}(k) = 1 - (1 - p)^k$. 
We compute $p$ from the observed data using~\Cref{eq:watts-actual} as $p = P(1) = \tfrac{m(1)}{n(1)}$. This is the fraction of inactive authors with \textit{exactly} one contact who became active as $P_\text{base}(1) = 1 - (1 - p)^1 = p$. Like before, we calculate the cumulative target activation probability for the baseline $C_{\text{base}}(k)$ with $k$ or more contacts as

\begin{equation}
    C_{\text{base}}(k) = \frac{\sum_k^\infty P_\text{base}(k) \cdot n(k)}{\sum_k^\infty n(k)}.
    \label{eq:watts-baseline}
\end{equation}
The denominator is the same as in~\Cref{eq:watts-actual} and comes from the observed data. 
The numerator represents the expected number of active authors if the contacts affect the activation independently.

\subsection{Source activation probability} \label{sec:source-active} 
% In Experiment II, we investigate whether the exclusive inactive coauthors of a prominent scholar concerning productivity or impact on a given topic $t$ have a higher probability to start working on $t$ than those of a less prominent scholar. 
Let $n_a$ be the number of exclusive inactive coauthors of an active author $a$ in the IW. 
Let $m_a$ be the number of those exclusive inactive coauthors who become active in the AW. The \textit{source activation probability} of scholar $a$ is thus

\begin{equation}
    P_s^a=\frac{m_a}{n_a}.
    \label{eq:s_act_prob}
\end{equation}

We stress that, for the probability to be well-defined, $n_a$ must be greater than zero. Therefore, in our calculations, we focused on active authors with at least one exclusive inactive coauthor.

For any $0 \leq f \leq 1$, we compute the fraction $C_s(f)$ of all active authors whose source activation probability is greater than or equal to $f$. 
$C_s(f)$ is the complementary cumulative probability distribution of the source activation probability $P_s^a$. 
As expected, $C_s(f)$ quickly decreases to 0 with increasing $f$. %, and becomes noisy when $f$ gets close to $1$ because of weak statistics.  
Because the curves corresponding to two sets of active authors are effectively indistinguishable at the tail, we compare a pair of points at some threshold $f^\ast$. 
We call $C_s(f^\ast)$ the \textit{cumulative source activation}.

The choice of the threshold $f^\ast$ is important.  
Setting it to 0 or 1 would return the same probability for both sets of authors. 
It should not also be too small for numerical reasons. %, if the number of coauthors is small, the smallest non-zero fraction that can get activated cannot be too small. 
For example, if there are only five inactive coauthors, the smallest non-zero fraction cannot be smaller than $1/5=0.20$. 
Choosing too high a value instead would lead to weaker statistics. 
% Therefore, the density of non-zero activation fractions close to zero is low for numerical reasons.
So, we fix the value at 0.10 for the results in the main text (\Cref{fig:EXP2_10,fig:EXP_2C}) and report the results for 0.20 in SI \Cref{fig:supp-exp2ab-20,fig:supp-exp2c-20}.
% So, we should use a small threshold but not too close to zero to avoid unfair comparisons. So we picked $f^\ast=0.10$ for the main text and $f^\ast = 0.20$ for the SI. In the SI, we report the results for $f^\star=0.2$, which show that our conclusions hold. 

\subsection{Chaperoning propensity} \label{sec:chaperone}
Let $m_a$ be the number of exclusive inactive coauthors of an active author $a$ who become active in the AW, which is the same as the numerator of \Cref{eq:s_act_prob}. 
Let $i_a$ be the number of those authors who write their first paper on topic $t$ with $a$ in the AW.
The \textit{chaperoning probability} of $a$ is defined as 
\begin{equation}
    P_c^a = \frac{i_a}{m_a}.
\end{equation}
 
We define the \textit{chaperoning propensity} $P_c(f)$ corresponding to a specific threshold $f \in [0, 1]$ as the fraction of all active authors with $P_c^a \ge f$.
We use the aforementioned values of $0.10$ (\Cref{fig:EXP2_10,fig:EXP_2C}) and $0.20$ (SI \Cref{fig:supp-exp2ab-20,fig:supp-exp2c-20}) for the threshold $f$. 

\begin{acknowledgments}
    This project was supported by grants from the National Science Foundation (\#1927418)
    and the Air Force Office of Scientific Research (\#FA9550-19-1-0354). This research was supported in part by Lilly Endowment, Inc., through its support for the Indiana University Pervasive Technology Institute.
\end{acknowledgments}

% \afterpage{\clearpage}
\appendix
\setcounter{totalnumber}{6}  % allow 6 floats per page 
\setcounter{topnumber}{3}
\setcounter{bottomnumber}{3}
\setcounter{dbltopnumber}{6}

\renewcommand{\thetable}{S\arabic{table}}   
\renewcommand{\thefigure}{S\arabic{figure}} 
\setcounter{figure}{0}
\setcounter{table}{0}

\clearpage 
\begin{table}
    \centering 
    \caption{Summary information for Physics topics. \#Papers: average number of Papers. \#Authors: average number of active authors. Averages are computed over all time windows selected for a topic.}
    \label{tab:Physics-info}
    \begin{tabular}{@{}lcrrrr@{}}
        \toprule
        \multirow{2}{*}{\textbf{Topic}} & \multirow{2}{*}{\textbf{\#Windows}} & \multicolumn{2}{c}{\textbf{Interaction Window}} & \multicolumn{2}{c}{\textbf{Activation Window}} \\
        \cmidrule(lr){3-4} \cmidrule{5-6} 
         & & \#Papers & \#Authors & \#Papers & \#Authors  \\
        \midrule
        Gravitational Wave & 10 & 3,613.70 & 5,745.20 & 5,486.40 & 9,160.30 \\
        Dark Matter & 13 & 6,433.69 & 8,348.23 & 9,203.38 & 12,346.00 \\
        Fluid Dynamics & 16 & 5,290.75 & 11,950.38 & 7,231.25 & 16,960.50 \\
        Soliton & 18 & 4,004.39 & 5,715.61 & 4,700.89 & 7,014.89 \\
        Supersymmetry & 20 & 5,328.85 & 4,827.45 & 5,470.75 & 5,361.25 \\
        Statistical Physics & 23 & 88,147.52 & 109,702.70 & 105,018.87 & 137,680.65 \\
        Superconductivity & 23 & 24,038.35 & 33,606.04 & 23,218.52 & 34,874.74 \\
        \bottomrule
    \end{tabular}

    \caption{Summary information for Computer Science topics. \#Papers: average number of Papers. \#Authors: average number of active authors. Averages are computed over all time windows selected for a topic.}
    \label{tab:CS-info}
    \begin{tabular}{@{}lcrrrr@{}}
        \toprule
        \multirow{2}{*}{\textbf{Topic}} & \multirow{2}{*}{\textbf{\#Windows}} & \multicolumn{2}{c}{\textbf{Interaction Window}} & \multicolumn{2}{c}{\textbf{Activation Window}} \\
        \cmidrule(lr){3-4} \cmidrule{5-6} 
         & & \#Papers & \#Authors & \#Papers & \#Authors  \\
        \midrule
        Compiler & 13 & 3,786.31 & 7,869.23 & 4,208.46 & 9,701.92 \\
        Mobile Computing & 13 & 6,356.00 & 13,844.77 & 6,828.77 & 15,827.85 \\
        Cryptography & 15 & 9,706.47 & 15,181.93 & 14,865.13 & 25,218.93 \\
        Cluster Analysis & 21 & 18,585.57 & 36,645.95 & 30,996.52 & 63,910.10 \\
        Image Processing & 23 & 13,149.65 & 28,191.35 & 16,617.65 & 38,089.70 \\
        Parallel Computing & 23 & 31,453.30 & 48,006.87 & 38,271.61 & 61,960.22 \\
        \bottomrule
    \end{tabular}

    \caption{Summary information for Biology \& Medicine topics. \#Papers: average number of Papers. \#Authors: average number of active authors. Averages are computed over all time windows selected for a topic.}
    \label{tab:BioMed-info}
    \begin{tabular}{@{}lcrrrr@{}}
        \toprule
        \multirow{2}{*}{\textbf{Topic}} & \multirow{2}{*}{\textbf{\#Windows}} & \multicolumn{2}{c}{\textbf{Interaction Window}} & \multicolumn{2}{c}{\textbf{Activation Window}} \\
        \cmidrule(lr){3-4} \cmidrule{5-6} 
         & & \#Papers & \#Authors & \#Papers & \#Authors  \\
        \midrule
        Protein Structure & 19 & 6,379.95 & 17,583.68 & 7,149.11 & 20,967.63 \\
        Genome & 23 & 28,066.09 & 71,481.87 & 44,089.78 & 114,696.48 \\
        Peptide Sequence & 23 & 12,347.48 & 43,348.96 & 9,733.04 & 37,330.09 \\
        Alzheimer's Disease & 23 & 9,313.78 & 22,628.30 & 11,723.22 & 31,624.61 \\
        Neurology & 23 & 9,260.17 & 26,046.57 & 12,795.70 & 39,515.00 \\
        Chemotherapy & 23 & 36,280.48 & 104,649.39 & 47,760.09 & 143,505.65 \\
        Radiation Therapy & 23 & 30,926.39 & 76,314.48 & 43,963.57 & 110,397.96 \\
        \bottomrule
    \end{tabular}
\end{table}

% \newpage 
\begin{table}
    \centering 
    \caption{ 
    Physics average Overlap Coefficient between the top and the bottom 10\% of active authors selected based on Productivity and two different definitions of Impact. The first definition uses $C_\text{avg}$ and is used in the main text. The second definition uses $C_\text{tot}$. The degree of overlap is significantly greater for $C_\text{tot}$.}
    \label{tab:Physics-overlap}
    \begin{tabular}{@{} lrrrr @{}}
        \toprule
        \multirow{2}{*}{\textbf{Topic}} & \multicolumn{2}{c}{\textbf{Top 10\%}} & \multicolumn{2}{c}{\textbf{Bottom 10\%}} \\
         \cmidrule(lr){2-3} \cmidrule{4-5}
         &  $C_\text{avg}$ &  $C_\text{tot}$ & $C_\text{avg}$ &  $C_\text{tot}$ \\
        \midrule
        Gravitational Wave & 0.33 & 0.59 & 0.14 & 0.14 \\
        Dark Matter & 0.31 & 0.56 & 0.15 & 0.15 \\
        Fluid Dynamics & 0.24 & 0.38 & 0.11 & 0.11 \\
        Soliton & 0.30 & 0.54 & 0.14 & 0.13 \\
        Supersymmetry & 0.30 & 0.58 & 0.17 & 0.16 \\
        Statistical Physics & 0.32 & 0.56 & 0.13 & 0.13 \\
        Superconductivity & 0.26 & 0.60 & 0.16 & 0.15 \\
        \bottomrule
    \end{tabular}
    
    \addtabletext{$C_\text{avg}$: Average of incoming citations from papers on the topic. $C_\text{tot}$: Sum of incoming citations from papers on the topic over all windows.}

    \caption{Computer Science average Overlap Coefficient between the top and the bottom 10\% of active authors selected based on Productivity and two different definitions of Impact. The first definition uses $C_\text{avg}$ and is used in the main text. The second definition uses $C_\text{tot}$. The degree of overlap is significantly greater for $C_\text{tot}$.}
    \label{tab:CS-overlap}
    \begin{tabular}{@{} lrrrr @{}}
        \toprule
        \multirow{2}{*}{\textbf{Topic}} & \multicolumn{2}{c}{\textbf{Top 10\%}} & \multicolumn{2}{c}{\textbf{Bottom 10\%}} \\
         \cmidrule(lr){2-3} \cmidrule{4-5}
         &  $C_\text{avg}$ &  $C_\text{tot}$ &  $C_\text{avg}$ &  $C_\text{tot}$ \\
        \midrule
        Compiler & 0.27 & 0.46 & 0.12 & 0.11 \\
        Mobile Computing & 0.25 & 0.41 & 0.12 & 0.12 \\
        Cryptography & 0.28 & 0.51 & 0.12 & 0.12 \\
        Cluster Analysis & 0.25 & 0.41 & 0.12 & 0.12 \\
        Image Processing & 0.25 & 0.42 & 0.12 & 0.11 \\
        Parallel Computing & 0.24 & 0.53 & 0.13 & 0.13 \\
        \bottomrule
    \end{tabular}

    \addtabletext{$C_\text{avg}$: Average of incoming citations from papers on the topic. $C_\text{tot}$: Sum of incoming citations from papers on the topic over all windows.}

    \caption{Biology \& Medicine average Overlap Coefficient between the top and the bottom 10\% of active authors selected based on Productivity and two different definitions of Impact. The first definition uses $C_\text{avg}$ and is used in the main text. The second definition uses $C_\text{tot}$. The degree of overlap is significantly greater for $C_\text{tot}$.}
    \label{tab:BioMed-overlap}
    \begin{tabular}{@{} lrrrr @{}}
        \toprule
        \multirow{2}{*}{\textbf{Topic}} & \multicolumn{2}{c}{\textbf{Top 10\%}} & \multicolumn{2}{c}{\textbf{Bottom 10\%}} \\
         \cmidrule(lr){2-3} \cmidrule{4-5}
         &  $C_\text{avg}$ &  $C_\text{tot}$ &  $C_\text{avg}$ &  $C_\text{tot}$ \\
        \midrule
        Protein Structure & 0.22 & 0.46 & 0.13 & 0.13 \\
        Genome & 0.22 & 0.50 & 0.13 & 0.13 \\
        Peptide Sequence & 0.18 & 0.41 & 0.12 & 0.12 \\
        Alzheimer's Disease & 0.19 & 0.55 & 0.13 & 0.14 \\
        Neurology & 0.16 & 0.37 & 0.12 & 0.12 \\
        Chemotherapy & 0.22 & 0.54 & 0.13 & 0.13 \\
        Radiation Therapy & 0.24 & 0.54 & 0.13 & 0.13 \\
        \bottomrule
    \end{tabular}
    
    \addtabletext{$C_\text{avg}$: Average of incoming citations from papers on the topic. $C_\text{tot}$: Sum of incoming citations from papers on the topic over all windows.}
\end{table}
 
% \afterpage{\clearpage}

\begin{figure*}
    \centering 
    \input{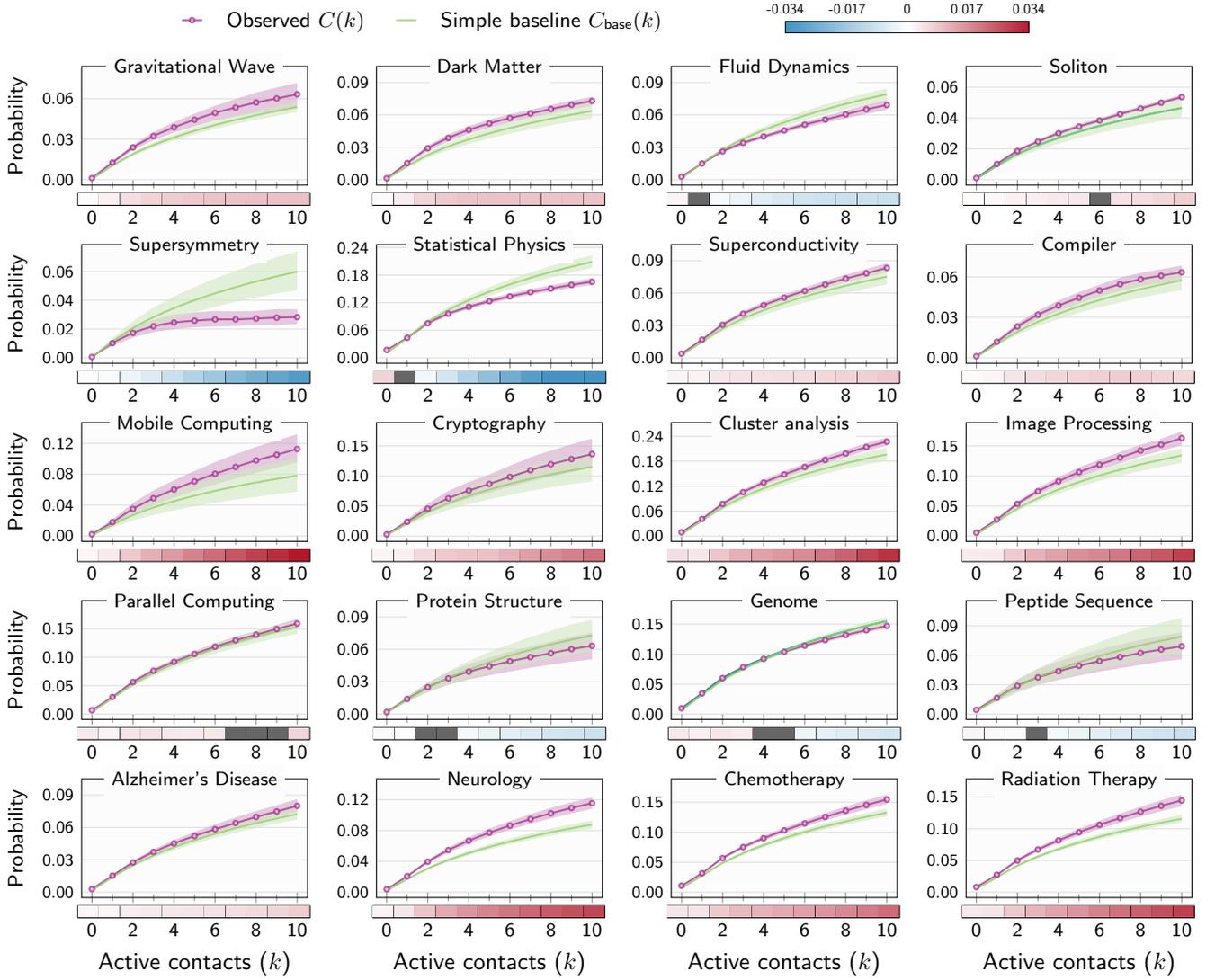}
    \caption{Experiment I. Same as \Cref{fig:membership-closure}, but here contacts is the number of papers written with active coauthors in the IW. \\
    Cumulative target activation probability (in purple) for inactive authors in the AW with shaded 95\% confidence intervals. For each $k$, the $y$-value indicates the fraction of inactive authors with at least $k$ active contacts in the IW who became active in the AW.
    The green solid line with shaded errors represents the baseline described in the text, corresponding to independent effects from the coauthors.
    The heatmap below the $x$-axis shows the mean difference between the observed and baseline curves for each $k$-value. It is gray if the 95\% confidence interval contains 0, denoting the $k$-values where the points are statistically indistinguishable at $p$-value $0.05$. Positive and negative deviations from the baseline are in red and blue, respectively.}
    \label{fig:supp-closure}
\end{figure*}
\newpage

\begin{figure}
    \centering 
    \input{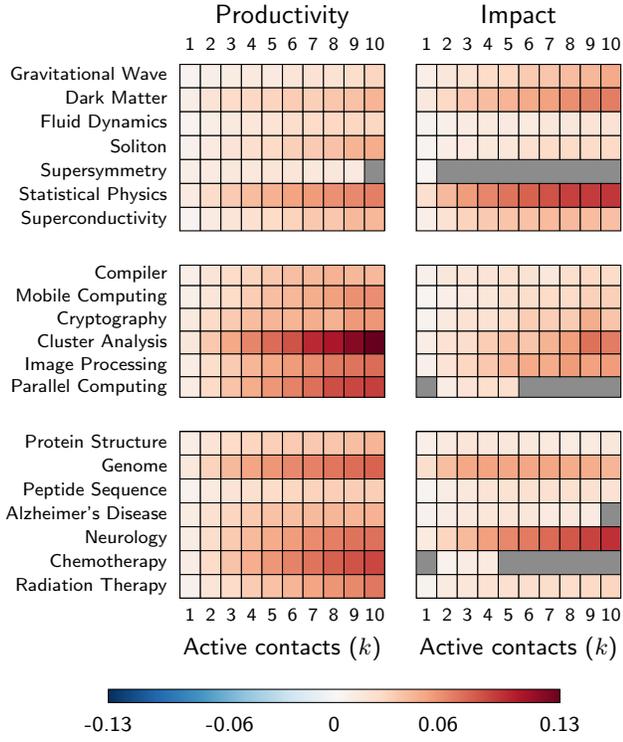}

% \pgfplotsset{colormap/PuOr}
\pgfplotsset{colormap/RdBu}
% \pgfplotsset{colormap/temp}

\begin{tikzpicture}
	\begin{groupplot}[
		group style={
			group size=2 by 3,
            group name=mygroup,
% 			vertical sep=12pt,
			vertical sep=13pt,
            horizontal sep=12pt,
			% x descriptions at=edge bottom,
            xlabels at=edge top,
            xticklabels at=edge bottom,
			y descriptions at=edge left,
		},
		width=0.24\textwidth,
        height=108pt,
		enlarge x limits={rel=0.05, upper},
		enlarge y limits={rel=0.08},
        typeset ticklabels with strut,
        title style={yshift=0.5em, align=center},
		colormap name=RdBu,
		ytick=data,
        % colorbar,
        tick align=outside,
		xtick=data,
        xtick style={draw=none},
		ytick style={draw=none, font=\scriptsize\sffamily, xshift=-5},
		xticklabel style={draw=none, font=\scriptsize\sffamily},
        yticklabel style={draw=none, font=\scriptsize\sffamily},
		axis background/.style={fill=gray!90}
	]
	%%%%%%%%% Physics
	\nextgroupplot[
		title={Productivity},
		ylabel={},
        point meta min=-0.13,
		point meta max=0.13,
        symbolic x coords={1,2,3,4,5,6,7,8,9,10},
        symbolic y coords={Gravitational wave,Dark matter,Fluid dynamics,Soliton,Supersymmetry,Statistical physics,Superconductivity},
        yticklabels={Gravitational Wave,Dark Matter,Fluid Dynamics,Soliton,Supersymmetry,Statistical Physics,Superconductivity},
        xticklabel pos=top,
        xticklabel style={yshift=-0.5em},
        yticklabel style={xshift=0.3em},
	]
	\addplot[
		matrix plot,  % not matrix plot* 
		draw=black, 
		fill opacity=1,
		colorbar source,
		 % nodes near coords,%=\coordindex,
		% mark=none,
		mesh/cols=10,
		point meta=explicit,
	] 
	table[x=k, y=topic, meta expr=-\thisrow{value}] {\PhysicsHeatmapProdVerTwo};	
	
    %%%%%%%%%
	\nextgroupplot[
		title={Impact},
		ylabel={},
        point meta min=-0.13,
		point meta max=0.13,
        symbolic x coords={1,2,3,4,5,6,7,8,9,10},
		symbolic y coords={Gravitational wave,Dark matter,Fluid dynamics,Soliton,Supersymmetry,Statistical physics,Superconductivity},
        yticklabels={},	
        xticklabel pos=top,
        xticklabel style={yshift=-0.5em},
	]
	\addplot[
		matrix plot,
		draw=black, 
		fill opacity=1,
		colorbar source,
		mesh/cols=10,
		point meta=explicit,
	] 
	table[x=k, y=topic, meta expr=-\thisrow{value}] {\PhysicsHeatmapImpOneVerTwo};	

    %%%%%%%%% CS
	\nextgroupplot[
		height=95pt,
        title={},
		ylabel={},
        point meta min=-0.13,
		point meta max=0.13,
        % colorbar,
		symbolic x coords={1,2,3,4,5,6,7,8,9,10},
		symbolic y coords={Compiler,Mobile computing,Cryptography,Cluster analysis,Image processing,Parallel computing},
        xticklabels={},       
        yticklabels={Compiler, Mobile Computing, Cryptography, Cluster Analysis, Image Processing, Parallel Computing},
        yticklabel style={xshift=0.3em},
	]
	\addplot[
		matrix plot,
		draw=black, 
		fill opacity=1,
		colorbar source,
		mesh/cols=10,
		point meta=explicit,
	] 
	table[x=k, y=topic, meta expr=-\thisrow{value}] {\CSHeatmapProdVerTwo};	
	
    %%%%%%%%%
	\nextgroupplot[
        height=95pt,
		title={},
		ylabel={},
        point meta min=-0.13,
		point meta max=0.13,
        % colorbar,
		symbolic x coords={1,2,3,4,5,6,7,8,9,10},
		symbolic y coords={Compiler,Mobile computing,Cryptography,Cluster analysis,Image processing,Parallel computing},
		% xticklabel pos=top,
        xticklabels={},
        yticklabels={},
	]
	\addplot[
		matrix plot,
		draw=black, 
		fill opacity=1,
		colorbar source,
		mesh/cols=10,
		point meta=explicit,
	] 
	table[x=k, y=topic, meta expr=-\thisrow{value}] {\CSHeatmapImpOneVerTwo};

 %%%%%%%%% BioMed
	\nextgroupplot[
		title={},
		ylabel={},
        point meta min=-0.13,
		point meta max=0.13,
        % colorbar,
		symbolic x coords={1,2,3,4,5,6,7,8,9,10},
		symbolic y coords={Protein structure,Genome,Peptide sequence,Alzheimer's disease,Neurology,Chemotherapy,Radiation therapy},
        yticklabels={Protein Structure, Genome, Peptide Sequence, Alzheimer's Disease, Neurology, Chemotherapy, Radiation Therapy},
        xticklabel style={yshift=0.5em},
        yticklabel style={xshift=0.3em},
		% xticklabel pos=top,
        % xticklabels={},       
	]
	\addplot[
		matrix plot,
		draw=black, 
		fill opacity=1,
		colorbar source,
		mesh/cols=10,
		point meta=explicit,
	] 
	table[x=k, y=topic, meta expr=-\thisrow{value}] {\BioMedHeatmapProdVerTwo};	
	
    %%%%%%%%%
	\nextgroupplot[
		title={},
		ylabel={},
        point meta min=-0.13,
		point meta max=0.13,
        % colorbar,
		% symbolic x coords={1,2,3,4,5,6,7,8,9,10},
		symbolic y coords={Protein structure,Genome,Peptide sequence,Alzheimer's disease,Neurology,Chemotherapy,Radiation therapy},
		% xticklabel pos=top,
        % xticklabels={},
        yticklabels={},
        xticklabel style={yshift=0.5em},
	]
	\addplot[
		matrix plot,
		draw=black, 
		fill opacity=1,
		colorbar source,
		mesh/cols=10,
		point meta=explicit,
	] 
	table[x=k, y=topic, meta expr=-\thisrow{value}] {\BioMedHeatmapImpOneVerTwo};

	\end{groupplot}

    \node[anchor=south, align=center] at ($(mygroup c1r3.south)+(0em, -2.9em)$) {Active contacts ($k$)};

    \node[anchor=south, align=center] at ($(mygroup c2r3.south)+(0em, -2.9em)$) {Active contacts ($k$)};

\begin{scope}[shift={(-0.95, -5.8)}, scale=1]
    \pgfplotscolorbardrawstandalone[
        colorbar horizontal,
        point meta min=-0.13, point meta max=0.13,
        colormap name=RdBu,
        colorbar style={
            x dir=reverse,
            height=5pt,
	    	xtick={-0.13,-0.06,0,0.06,0.13}, 
            xticklabels={0.13,0.06,0,-0.06,-0.13},  % need to manually 
            scale=1,
            xtick style={
                draw=none,
                /pgf/number format/.cd,
                fixed,
                fixed zerofill,
                precision=0,
                /tikz/.cd
        	},
            xticklabel style={font=\footnotesize\sffamily},
            xticklabel pos=lower,
            scaled ticks=false,
        },
    ]
\end{scope}
\end{tikzpicture}
    \caption{Experiment I. Same as \Cref{fig:closure-heatmaps}, but here contacts is the number of papers written with active coauthors in the IW. 
    Heatmaps showing the mean difference between the cumulative target activation probabilities of the inactive authors in the AW who had exclusive contacts with the top 10\% and bottom 10\% of active authors, respectively, selected according to productivity (left) and impact (right) in the IW. The cells are gray if the 95\% confidence interval contains 0. The majority of red cells indicate that the cumulative target activation probabilities for contacts with the top 10\% are higher than those with the bottom 10\%. }
    \label{fig:supp-heatmaps}
\end{figure}

% \afterpage{\clearpage}
% \newpage 

\begin{figure}
    \centering 
    \input{data/influence}

\begin{tikzpicture}

\begin{scope}[shift={(4, 4.1)}]
    \begin{customlegend}[ 
    legend columns=4,
    legend style={
    draw=none,
    column sep=2ex,
    % nodes={scale=1.8, transform shape},
  },
  legend cell align={left},
  legend entries={Productivity~~,Impact}
  ]
    \addlegendimage{Accent-A, mark options={draw=Accent-A!70!black, thick, fill=Accent-A!70!white,}, draw opacity=0.8, only marks,}
    \addlegendimage{PiYG-E, mark options={draw=PiYG-E!70!black, thick, fill=PiYG-E!70!white}, draw opacity=0.8, only marks,}
    \end{customlegend}
\end{scope}

\begin{groupplot}[
        group style={
            group name=my plots,
            % group size=3 by 3,
            rows=3, columns=2,
            % x descriptions at=edge bottom,
            xticklabels at=all,
            xlabels at=edge bottom,
            y descriptions at=edge left,
            vertical sep=25pt,
            horizontal sep=20pt,
        },
        height=150pt,
        width=130pt,
        xmajorgrids,
        % ymajorgrids,
        major y grid style={draw=gray, draw opacity=0.4},
        ytick style={yshift=0pt},
        grid style={draw=gray!50, draw opacity=0.5},
        % xtick align=center, 
        % xtick style={draw=none},
        ytick style={draw=none},
        scaled x ticks=false,
        scaled y ticks=false,
        yticklabel style={yshift=-5pt, xshift=-12pt, font=\scriptsize\sffamily},
        axis y line*=middle,
        axis x line*=bottom, 
        enlarge y limits={0.05, upper},
        enlarge y limits={0.15, lower},
        enlarge x limits=0.05,
        title style={yshift=0em, align=center},
        xlabel={Difference}, ylabel={}, 
        axis background/.style={fill=gray!3},
    ]

%%%% Physics - Exp A
    
    \nextgroupplot[
        title={A}, 
        symbolic y coords={Superconductivity,Statistical physics,Supersymmetry,Soliton,Fluid dynamics,Dark matter,Gravitational wave},
        ytick={Superconductivity,Statistical physics,Supersymmetry,Soliton,Fluid dynamics,Dark matter,Gravitational wave},
        yticklabels={Superconductivity, Statistical Physics, Supersymmetry, Soliton, Fluid Dynamics, Dark Matter, Gravitational Wave},
        xmin=-0.025, xmax=0.07, 
        xtick={-0.025,0,0.03,0.06}, xticklabels={-0.025,0,0.03,0.06},
        % minor xtick={0.025,0.075},
    ]
    
        \addplot [error1, error bars/.cd, x dir=both, x explicit, error bar style=thick, error mark=|, error mark options={thick},] table[x=Prod_diffA_mean, y=topic, x error plus expr=\thisrow{Prod_diffA_mean_ci95_max}-\thisrow{Prod_diffA_mean}, x error minus expr=\thisrow{Prod_diffA_mean}-\thisrow{Prod_diffA_mean_ci95_min}] {\PhysicsInfluenceProdImpOneTwenty}; 
        %\addlegendentry{Productivity};

        \addplot [error2, error bars/.cd, x dir=both, x explicit, error bar style=thick, error mark=|, error mark options={thick},] table[x=Imp1_diffA_mean, y=topic, x error plus expr=\thisrow{Imp1_diffA_mean_ci95_max}-\thisrow{Imp1_diffA_mean}, x error minus expr=\thisrow{Imp1_diffA_mean}-\thisrow{Imp1_diffA_mean_ci95_min}] {\PhysicsInfluenceProdImpOneTwenty}; 
        %\addlegendentry{Impact};

        % \addplot[mark=none, draw opacity=0.5, draw=gray] coordinates {(0,current bounding box.south) (0,Statistical physics)};
        % \draw[add=2 and .5] (axis cs:0,Gravitational wave) -- (axis cs:0,Statistical physics);

    %%%%% Exp B

    \nextgroupplot[
        title={B}, 
        symbolic y coords={Superconductivity,Statistical physics,Supersymmetry,Soliton,Fluid dynamics,Dark matter,Gravitational wave}, 
        ytick={Superconductivity,Statistical physics,Supersymmetry,Soliton,Fluid dynamics,Dark matter,Gravitational wave},
        yticklabels=\empty,
        xmin=-0.05, xmax=0.3, 
        xtick={-0.05,0,0.1,0.2,0.3}, xticklabels={-0.05,0,0.1,0.2,0.3},
        minor xtick={0.05,0.15,0.25},
    ]
        \addplot [error1, error bars/.cd, x dir=both, x explicit, error bar style=thick, error mark=|, error mark options={thick},] table[x=Prod_diffB_mean, y=topic, x error plus expr=\thisrow{Prod_diffB_mean_ci95_max}-\thisrow{Prod_diffB_mean}, x error minus expr=\thisrow{Prod_diffB_mean}-\thisrow{Prod_diffB_mean_ci95_min}] {\PhysicsInfluenceProdImpOneTwenty};         

        \addplot [error2, error bars/.cd, x dir=both, x explicit, error bar style=thick, error mark=|, error mark options={thick},] table[x=Imp1_diffB_mean, y=topic, x error plus expr=\thisrow{Imp1_diffB_mean_ci95_max}-\thisrow{Imp1_diffB_mean}, x error minus expr=\thisrow{Imp1_diffB_mean}-\thisrow{Imp1_diffB_mean_ci95_min}] {\PhysicsInfluenceProdImpOneTwenty};

    %%%%%  CS Exp A

    \nextgroupplot[
        title={},
        symbolic y coords={Parallel computing,Image processing,Cluster analysis,Cryptography,Mobile computing,Compiler},
        ytick={Parallel computing,Image processing,Cluster analysis,Cryptography,Mobile computing,Compiler},
        yticklabels={Parallel Computing,Image Processing,Cluster Analysis,Cryptography,Mobile Computing,Compiler},
        xmin=-0.025, xmax=0.06, 
        xtick={-0.025,0,0.03,0.06}, xticklabels={-0.025,0,0.03,0.06},
        % minor xtick={0.025,0.075},
    ]

        \addplot [error1, error bars/.cd, x dir=both, x explicit, error bar style=thick, error mark=|, error mark options={thick},] table[x=Prod_diffA_mean, y=topic, x error plus expr=\thisrow{Prod_diffA_mean_ci95_max}-\thisrow{Prod_diffA_mean}, x error minus expr=\thisrow{Prod_diffA_mean}-\thisrow{Prod_diffA_mean_ci95_min}] {\CSInfluenceProdImpOneTwenty}; 

        \addplot [error2, error bars/.cd, x dir=both, x explicit, error bar style=thick, error mark=|, error mark options={thick},] table[x=Imp1_diffA_mean, y=topic, x error plus expr=\thisrow{Imp1_diffA_mean_ci95_max}-\thisrow{Imp1_diffA_mean}, x error minus expr=\thisrow{Imp1_diffA_mean}-\thisrow{Imp1_diffA_mean_ci95_min}] {\CSInfluenceProdImpOneTwenty};
        
    %%%%%  CS Exp B

    \nextgroupplot[
        title={},
        symbolic y coords={Parallel computing,Image processing,Cluster analysis,Cryptography,Mobile computing,Compiler}, 
        ytick={Parallel computing,Image processing,Cluster analysis,Cryptography,Mobile computing,Compiler},
        yticklabels=\empty,
        xmin=-0.05, xmax=0.3, 
        xtick={-0.05,0,0.1,0.2,0.3}, xticklabels={-0.05,0,0.1,0.2,0.3},
        minor xtick={0.05,0.15,0.25},
    ]
        
        \addplot [error1, error bars/.cd, x dir=both, x explicit, error bar style=thick, error mark=|, error mark options={thick},] table[x=Prod_diffB_mean, y=topic, x error plus expr=\thisrow{Prod_diffB_mean_ci95_max}-\thisrow{Prod_diffB_mean}, x error minus expr=\thisrow{Prod_diffB_mean}-\thisrow{Prod_diffB_mean_ci95_min}] {\CSInfluenceProdImpOneTwenty};
        
        \addplot [error2, error bars/.cd, x dir=both, x explicit, error bar style=thick, error mark=|, error mark options={thick},] table[x=Imp1_diffB_mean, y=topic, x error plus expr=\thisrow{Imp1_diffB_mean_ci95_max}-\thisrow{Imp1_diffB_mean}, x error minus expr=\thisrow{Imp1_diffB_mean}-\thisrow{Imp1_diffB_mean_ci95_min}] {\CSInfluenceProdImpOneTwenty};

    %%%% BioMed Exp A
    
    \nextgroupplot[
        title={}, 
        symbolic y coords={Radiation therapy,Chemotherapy,Neurology,Alzheimer's disease,Peptide sequence,Genome,Protein structure}, 
        ytick={Radiation therapy,Chemotherapy,Neurology,Alzheimer's disease,Peptide sequence,Genome,Protein structure},
        yticklabels={Radiation Therapy,Chemotherapy,Neurology,Alzheimer's Disease,Peptide Sequence,Genome,Protein Structure},
        xmin=-0.025, xmax=0.06, 
        xtick={-0.025,0,0.03,0.06}, xticklabels={-0.025,0,0.03,0.06},
        % minor xtick={0.025,0.075},
    ]
        
        \addplot [error1, error bars/.cd, x dir=both, x explicit, error bar style=thick, error mark=|, error mark options={thick},] table[x=Prod_diffA_mean, y=topic, x error plus expr=\thisrow{Prod_diffA_mean_ci95_max}-\thisrow{Prod_diffA_mean}, x error minus expr=\thisrow{Prod_diffA_mean}-\thisrow{Prod_diffA_mean_ci95_min}] {\BioMedInfluenceProdImpOneTwenty}; 

        \addplot [error2, error bars/.cd, x dir=both, x explicit, error bar style=thick, error mark=|, error mark options={thick},] table[x=Imp1_diffA_mean, y=topic, x error plus expr=\thisrow{Imp1_diffA_mean_ci95_max}-\thisrow{Imp1_diffA_mean}, x error minus expr=\thisrow{Imp1_diffA_mean}-\thisrow{Imp1_diffA_mean_ci95_min}] {\BioMedInfluenceProdImpOneTwenty}; 

    %%%%%%%%%% Exp B
    \nextgroupplot[
        title={},
        symbolic y coords={Radiation therapy,Chemotherapy,Neurology,Alzheimer's disease,Peptide sequence,Genome,Protein structure}, 
        ytick={Radiation therapy,Chemotherapy,Neurology,Alzheimer's disease,Peptide sequence,Genome,Protein structure},
        yticklabels=\empty,  
        xmin=-0.05, xmax=0.3, 
        xtick={-0.05,0,0.1,0.2,0.3}, xticklabels={-0.05,0,0.1,0.2,0.3},
        minor xtick={0.05,0.15,0.25},
    ]

        \addplot [error1, error bars/.cd, x dir=both, x explicit, error bar style=thick, error mark=|, error mark options={thick},] table[x=Prod_diffB_mean, y=topic, x error plus expr=\thisrow{Prod_diffB_mean_ci95_max}-\thisrow{Prod_diffB_mean}, x error minus expr=\thisrow{Prod_diffB_mean}-\thisrow{Prod_diffB_mean_ci95_min}] {\BioMedInfluenceProdImpOneTwenty};

        \addplot [error2, error bars/.cd, x dir=both, x explicit, error bar style=thick, error mark=|, error mark options={thick},] table[x=Imp1_diffB_mean, y=topic, x error plus expr=\thisrow{Imp1_diffB_mean_ci95_max}-\thisrow{Imp1_diffB_mean}, x error minus expr=\thisrow{Imp1_diffB_mean}-\thisrow{Imp1_diffB_mean_ci95_min}] {\BioMedInfluenceProdImpOneTwenty};

\end{groupplot}
\end{tikzpicture}
    \caption{Experiment II. Same as \Cref{fig:EXP2_10}, with threshold $f^\star = 0.20$. (A) Mean and 95\% confidence interval of the means of the difference between the cumulative source activations of active authors in the top 10\% and bottom 10\% based on productivity (green) and impact (pink). (B) Mean and 95\% confidence interval of the means of the difference between the cumulative chaperoning propensities of active authors in the top 10\% and bottom 10\% based on productivity (green) and impact (pink). A positive difference indicates that the effect is stronger for the top 10\% active authors.}
    \label{fig:supp-exp2ab-20}
\end{figure}
% \clearpage 

\begin{figure*}
    \centering 
    \input{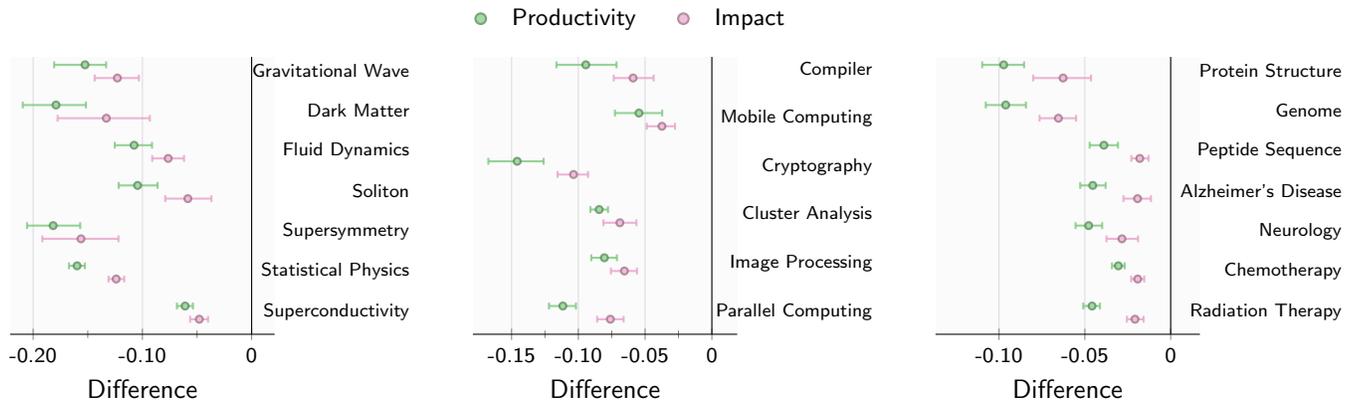}

\begin{tikzpicture}

\begin{scope}[shift={(9.5, 3.5)}]
    \begin{customlegend}[ 
    legend columns=2,
    legend style={
    draw=none,
    column sep=2ex,
    % nodes={scale=1.8, transform shape},
  },
  legend cell align={left},
  legend entries={Productivity~~~,Impact}
  ]
    \addlegendimage{Accent-A, mark options={draw=Accent-A!70!black, thick, fill=Accent-A!70!white,}, draw opacity=0.8, only marks,}
    \addlegendimage{PiYG-E, mark options={draw=PiYG-E!70!black, thick, fill=PiYG-E!70!white}, draw opacity=0.8, only marks,}
    \end{customlegend}
\end{scope}

\begin{groupplot}[
        group style={
            group name=my plots,
            % group size=3 by 3,
            rows=1, columns=3,
            % x descriptions at=edge bottom,
            xticklabels at=all,
            xlabels at=edge bottom,
            y descriptions at=edge left,
            vertical sep=25pt,
            % horizontal sep=15pt,
            horizontal sep=75pt,
        },
        height=150pt,
        width=145pt,
        xmajorgrids,
        % ymajorgrids,
        major y grid style={draw=gray, draw opacity=0.4},
        ytick style={yshift=0pt},
        grid style={draw=gray!50, draw opacity=0.5},
        % xtick align=center, 
        % xtick style={draw=none},
        ytick style={draw=none},
        yticklabel style={yshift=-5pt, xshift=68pt, font=\scriptsize\sffamily},
        % axis y line*=none,
        axis x line*=bottom,
        axis y line*=middle,
        % every inner x axis line/.append style={--},
        % y axis line style={draw=none},
        % enlarge y limits={0.01, upper},
        enlarge y limits={0.05, upper},
        enlarge y limits={0.15, lower},
        enlarge x limits=0.05, 
        title style={yshift=0em, align=center},
        xlabel={Difference}, ylabel={}, 
        axis background/.style={fill=gray!3},
    ]
    %%%% Physics Exp C 
    
    \nextgroupplot[
        title={}, 
        yticklabel style={xshift=-5pt},
        symbolic y coords={Superconductivity,Statistical physics,Supersymmetry,Soliton,Fluid dynamics,Dark matter,Gravitational wave}, 
        ytick={Superconductivity,Statistical physics,Supersymmetry,Soliton,Fluid dynamics,Dark matter,Gravitational wave},
        yticklabels={Superconductivity, Statistical Physics, Supersymmetry, Soliton, Fluid Dynamics, Dark Matter, Gravitational Wave},  
        xmin=-0.21, xmax=0.01, 
        minor x tick num=1,
        xtick={-0.2,-0.1,0}, xticklabels={-0.20,-0.10,0},
        % minor xtick={0.05,0.15,0.25},
    ]
    
        \addplot [error1, error bars/.cd, x dir=both, x explicit, error bar style=thick, error mark=|, error mark options={thick},] table[x=Prod_diffC_mean, y=topic, x error plus expr=\thisrow{Prod_diffC_mean_ci95_max}-\thisrow{Prod_diffC_mean}, x error minus expr=\thisrow{Prod_diffC_mean}-\thisrow{Prod_diffC_mean_ci95_min}] {\PhysicsInfluenceProdImpOneTwenty};

        \addplot [error2, error bars/.cd, x dir=both, x explicit, error bar style=thick, error mark=|, error mark options={thick},] table[x=Imp1_diffC_mean, y=topic, x error plus expr=\thisrow{Imp1_diffC_mean_ci95_max}-\thisrow{Imp1_diffC_mean}, x error minus expr=\thisrow{Imp1_diffC_mean}-\thisrow{Imp1_diffC_mean_ci95_min}] {\PhysicsInfluenceProdImpOneTwenty};

    %%% CS Exp C
    \nextgroupplot[
        title={}, 
        yticklabel style={xshift=-4pt},
        symbolic y coords={Parallel computing,Image processing,Cluster analysis,Cryptography,Mobile computing,Compiler}, 
        ytick={Parallel computing,Image processing,Cluster analysis,Cryptography,Mobile computing,Compiler}, 
        yticklabels={Parallel Computing, Image Processing, Cluster Analysis, Cryptography, Mobile Computing, Compiler},
        xmin=-0.17, xmax=0.01, 
        minor x tick num=1,
        xtick={-0.15,-0.1,-0.05,0}, xticklabels={-0.15,-0.10,-0.05,0},
        % minor xtick={0.05,0.15,0.25},
    ]

        \addplot [error1, error bars/.cd, x dir=both, x explicit, error bar style=thick, error mark=|, error mark options={thick},] table[x=Prod_diffC_mean, y=topic, x error plus expr=\thisrow{Prod_diffC_mean_ci95_max}-\thisrow{Prod_diffC_mean}, x error minus expr=\thisrow{Prod_diffC_mean}-\thisrow{Prod_diffC_mean_ci95_min}] {\CSInfluenceProdImpOneTwenty};

        \addplot [error2, error bars/.cd, x dir=both, x explicit, error bar style=thick, error mark=|, error mark options={thick},] table[x=Imp1_diffC_mean, y=topic, x error plus expr=\thisrow{Imp1_diffC_mean_ci95_max}-\thisrow{Imp1_diffC_mean}, x error minus expr=\thisrow{Imp1_diffC_mean}-\thisrow{Imp1_diffC_mean_ci95_min}] {\CSInfluenceProdImpOneTwenty};

    % %%%%%%% BioMed Exp C
    \nextgroupplot[
        title={},
        symbolic y coords={Radiation therapy,Chemotherapy,Neurology,Alzheimer's disease,Peptide sequence,Genome,Protein structure},
        ytick={Radiation therapy,Chemotherapy,Neurology,Alzheimer's disease,Peptide sequence,Genome,Protein structure},
        yticklabels={Radiation Therapy, Chemotherapy, Neurology, Alzheimer's Disease, Peptide Sequence, Genome, Protein Structure},
        xmin=-0.13, xmax=0.01, 
        minor x tick num=0,
        xtick={-0.15,-0.1,-0.05,0}, xticklabels={-0.15,-0.10,-0.05,0},
        % minor xtick={0.05,0.15,0.25},
    ]

        \addplot [error1, error bars/.cd, x dir=both, x explicit, error bar style=thick, error mark=|, error mark options={thick},] table[x=Prod_diffC_mean, y=topic, x error plus expr=\thisrow{Prod_diffC_mean_ci95_max}-\thisrow{Prod_diffC_mean}, x error minus expr=\thisrow{Prod_diffC_mean}-\thisrow{Prod_diffC_mean_ci95_min}] {\BioMedInfluenceProdImpOneTwenty};

        \addplot [error2, error bars/.cd, x dir=both, x explicit, error bar style=thick, error mark=|, error mark options={thick},] table[x=Imp1_diffC_mean, y=topic, x error plus expr=\thisrow{Imp1_diffC_mean_ci95_max}-\thisrow{Imp1_diffC_mean}, x error minus expr=\thisrow{Imp1_diffC_mean}-\thisrow{Imp1_diffC_mean_ci95_min}] {\BioMedInfluenceProdImpOneTwenty};

\end{groupplot}
\end{tikzpicture}
    \caption{Experiment II. Same as \Cref{fig:EXP_2C}, with threshold $f^\star = 0.20$. Dilution effect. The mean and 95\% confidence interval of the mean of the difference between the cumulative source activations of active authors in the top 20\% and bottom 20\% bins, based on the average number of coauthors, from the set of top 10\% active authors in productivity (green) and impact (pink).
    A negative difference across the topics indicates a \textit{dilution} effect, wherein coauthors of prominent active scholars with less collaborators (on average) are more likely to switch topics. }
    \label{fig:supp-exp2c-20}
\end{figure*}

\end{document}